\documentclass{edp-jp4}
\usepackage{graphicx}
\usepackage{psfrag}
\usepackage{amssymb}
\begin{document}

\title{Low-dimensional trapped gases} 
\author{D.S. Petrov$^{*\,1,2}$}
\author{D.M. Gangardt$^{**\,3}$}
\author{G.V. Shlyapnikov$^{**\,1,2,}$ }
\address{FOM Institute AMOLF, Kruislaan 407, 1098
SJ Amsterdam, The Netherlands} 
\address{Russian Research Center Kurchatov Institute, Kurchatov
Square, 123182 Moscow, Russia}
\address{Laboratoire Kastler Brossel, Ecole Normale Sup\'erieure,
24 rue Lhomond, 75231, Paris, France} 
%
\maketitle
\begin{abstract} 
Recent developments in the physics of ultracold gases provide wide 
possibilities for reducing the dimensionality of space for magnetically or
optically trapped atoms. The goal of these lectures is to show 
that regimes of quantum degeneracy in two-dimensional (2D) and 
one-dimensional (1D) trapped gases are drastically different from those 
in three dimensions and to stimulate an interest in low-dimensional systems.
Attention is focused on the new physics appearing in currently studied 
low-dimensional trapped gases and related to finite-size and finite-temperature 
effects. 
  
\end{abstract}
%
\renewcommand{\thefootnote}{\fnsymbol{footnote}}

\footnotetext[1]{Present address: ITAMP, Harvard-Smitsonian Center for
Astrophysics, and Harvard-MIT Center for Ultracold Atoms, 
Department of Physics, Cambridge, Massachusetts 02138, USA}
\footnotetext[7]{Present address: Laboratoire de Physique Th\'eorique
et Mod\`eles Statistiques, Universit\'e Paris Sud, 91405, Orsay Cedex,
France, and Van der Waals-Zeeman Institute, University of Amsterdam, 
Valckenierstraat 65/67, 1018 XE Amsterdam, The Netherlands}

\tableofcontents


\section*{Introduction}
\label{sec:Introduction}

The subject of low-dimensional quantum gases has a long pre-history. 
The influence of dimensionality of the system of bosons on the presence  
and character of Bose-Einstein condensation (BEC) and superfluid phase  
transition has been a subject of extensive studies in the spatially  
homogeneous case. From a general point of view, the absence of a true 
condensate in 2D and 1D at finite temperatures follows from the Bogoliubov 
$k^{-2}$ theorem and originates from long-wave fluctuations of the phase 
(see, e.g., \cite{Popov}). This has been expounded by Mermin and Wagner 
\cite{Mermin} and by Hohenberg \cite{Hoh}, and formed a basis for later
investigations.

The earlier discussion of low-dimensional Bose gases was mostly academic as
there was no possible realization of such a system. Fast progress in
evaporative and optical cooling of trapped atoms and the observation of 
Bose-Einstein condensation (BEC) in trapped clouds of alkali atoms 
\cite{discovery1,discovery2,discovery3} stimulated a search for 
non-trivial trapping geometries. 
Present facilities allow one to tightly confine the motion of trapped  
particles in one (two) direction(s) to zero point oscillations.
Then, kinematically the gas is 2D (1D), and the difference from
purely 2D (1D) gases is only related to the value of the effective
interparticle interaction which now depends on the tight confinement. 

Recent experiments have already reached 2D and 1D regimes for trapped 
Bose gases and studied some of the quantum degenerate states.
These studies bring in new physics originating from a finite size of
the system, spatial inhomogeneity, and finite temperatures. 
The present lectures cover most important issues in the physics of 
2D and 1D trapped quantum gases: the nature of various quantum
degenerate states, the role of interaction between particles, and
the role of finite-temperature and finite-size effects.

\section{{\it Lecture 1}. BEC in ideal 2D and 1D gases}

We start with describing a cross-over to the BEC regime in ideal 2D and 1D Bose
gases with a finite number of particles. We will consider an equilibrium gas at
temperature $T$ in the grand canonical ensemble, where the chemical potential
$\mu$ is fixed and the number of particles $N$ is fluctuating. In the
thermodynamic limit  ($N\rightarrow\infty$) this is equivalent to the
description in the canonical ensemble (fixed $N$ and fluctuating $\mu$).

In any dimension and confining potential the gas is characterized by a set
of eigenenergies of an individual particle, $E_{\nu}$, with the index $\nu$
labeling quantum numbers of the particle eigenstates. 
The (average) total number of particles $N$ is then related
to the temperature and chemical potential by the
equation
\begin{equation}
\label{Nnu}
N=\sum_\nu N_{\nu}\!\!\left(\frac{E_{\nu}-\mu}{T}\right)\quad ,
\end{equation}
where $N_{\nu}(z)=1/(\exp z-1)$ are the equilibrium occupation numbers of the
eigenstates. We now demonstrate how Eq.(\ref{Nnu}) allows one to establish the
presence or absence of BEC in 2D and 1D ideal Bose gases.

\subsection{Uniform ideal gas}

In an infinitely large uniform gas the particle eigenstates are characterized
by the momentum ${\bf k}$ and the eigenenergy $E_k=\hbar^2k^2/2m$, where $m$
is the mass of a particle. Then Eq.(\ref{Nnu}) takes the form
\begin{equation}    \label{Nk}
N=\Omega\int\!\!\frac{d^dk}{(2\pi)^d}\,N_k\!\!\left(\frac{E_k-\mu}{T}\right)
\quad,
\end{equation}
with $d$ being the dimension of the system, and $\Omega$ the $d$-dimensional
volume. 

In the 2D case the integration in Eq.(\ref{Nk}) is straightforward and we
obtain
\begin{equation}      \label{mu2Dun}
\mu=T\ln\!\left[1-\exp(-n_2\Lambda_T^2)\right]<0
\quad ,
\end{equation}
where $\Lambda_T=(2\pi\hbar^2/mT)^{1/2}$ is the thermal de Broglie wavelength,
and $n_2$ is the 2D density. The quantity $n_2\Lambda_T^2$ is called the
degeneracy parameter and in 2D it can be written as $n_2\Lambda_T^2=T/T_d$,
where $T_d=2\pi\hbar^2n_2/m$ is the temperature of quantum degeneracy. In the
limit of a classical gas, $n_2\Lambda_T^2\ll 1$, Eq.(\ref{mu2Dun}) gives the
well-known result $\mu=T\ln(n_2\Lambda_T^2)$. For a strongly degenerate gas,
where $n_2\Lambda_T^2\gg 1$, we obtain $\mu=-T\exp(-n_2\Lambda_T^2)$.   
Unlike in the 3D case, the dependence $\mu(T)$ is analytical and it shows a
monotonic increase of the chemical potential with decreasing temperature up
to $T\rightarrow 0$. In the thermodynamic limit the population of the ground
state ($k=0$) remains microscopic. One thus can say that there is no BEC in a
finite-temperature ideal uniform 2D Bose gas.

The situation is similar for an infinite uniform 1D Bose gas, where the 
degeneracy parameter is $n_1\Lambda_T=(T/T_d)^{1/2}$ and the temperature of
quantum degeneracy is given by $T_d=2\pi\hbar^2n_1^2/m$, with $n_1$ being 
the 1D density. In the classical limit ($n_1\Lambda_T\ll 1$), and in the limit
of a strongly degenerate gas ($n_1\Lambda_T\gg 1$), Eq.(\ref{Nk}) gives 
\begin{eqnarray}   
&&\mu=T\ln(n_1\Lambda_T) \quad , \qquad n_1\Lambda_T\ll 1 \quad ;   \label{mu1Dun1} \\
&&\mu=-\frac{\pi T}{(n_1\Lambda_T)^2} \quad , \qquad n_1\Lambda_T\gg 1 \quad .\label{mu1Dun2}
\end{eqnarray}  
Again, the chemical potential monotonically decreases with temperature and
remains negative at any $T$, which indicates the absence of BEC. 

The absence of BEC in infinitely large uniform 2D and 1D Bose gases is a
striking difference from the 3D case. This difference originates from the
energy dependence of the density of states. The (energy) density of states is
$\rho(E)\propto E^{(d/2-1)}$, where $d$ is the dimension of the system, and
for the 3D gas it decreases with $E$. Therefore, at sufficiently low
temperatures it becomes impossible to thermally occupy the low energy states
while maintaining a constant chemical potential or density. As a result, a
macroscopic number of particles goes to the ground state ($k=0$), {\it i.e.}
one has the phenomenon of BEC. In 2D and 1D the density of states does not
decrease with $E$ and this phenomenon is absent.    

\subsection{Ideal gas in a harmonic trap}  

For 2D and 1D Bose gases in a harmonic confining potential the density of
states is $\rho(E)\propto E^{(d-1)}$ and the situation changes. The population
of the ground state ($E=0$) is
\begin{equation}     \label{N0}
N_0=\frac{1}{\exp(-\mu/T)-1} \quad .
\end{equation}
For a large but finite number of particles in a trap, $N_0$ can become
macroscopic (comparable with $N$) at a small but finite negative $\mu$.
One then speaks of a cross-over to the BEC regime.

We first discuss the BEC cross-over for the 2D Bose gas in a symmetric harmonic
confining potential $V({\bf r})=m\omega^2(x^2+y^2)/2$. In this case the particle
energy is $E_{\nu}=\hbar\omega(n_x+n_y)$, with quantum numbers $n_x,n_y$ being
non-negative integers. The density of states is then
$\rho(E)=E/(\hbar\omega)^2$, and the contribution of low-energy excited states
to the total number of particles is negligible. Therefore, separating out
the population of the ground state, one can replace the summation in
Eq.(\ref{Nnu}) by integration:
\begin{equation}           \label{Nint} 
N=N_0+\int_0^{\infty}\!\!\!dE\, \rho(E) N\!\!\left(\frac{E-\mu}{T}\right) \quad . 
\end{equation}  
Assuming a large population of the ground state, from Eq.(\ref{N0}) we obtain
$-\mu/T\approx 1/N_0\ll 1$, 
and the population of excited trap states proves to be
\begin{equation}           \label{terms}
\int_0^{\infty}\!\!\!dE\, \rho(E) N\!\!\left(\frac{E-\mu}{T}\right)
\approx\left(\frac{T}{\hbar\omega}\right)^2\left(\frac{\pi^2}{6}-\frac{1+\ln
N_0}{N_0}\right)\quad . 
\end{equation}
This allows us to write Eq.(\ref{Nint}) in the form
\begin{equation}
\label{GroundPopeq}
N\left[1-\left(\frac{T}{T_c}\right)^2\right]=N_0-
\left(\frac{T}{\hbar\omega}\right)^2\frac{1+\ln N_0}{N_0} \quad ,
\end{equation}
where
\begin{equation}\label{Tc2D}
T_c=\sqrt{\frac{6N}{\pi^2}}\hbar\omega\quad .
\end{equation}

For a large number of particles, Eq.~(\ref{GroundPopeq}) indicates the
presence of a sharp cross-over to the  BEC regime at $T\approx T_c$. Below
$T_c$ we omit the last term in Eq.(\ref{GroundPopeq}) and obtain the
occupation of the ground  state 
\begin{equation}    \label{N02D}
N_0\approx N\left[1-\left(\frac{T}{T_c}\right)^2\right] \quad . 
\end{equation}
Note that at $T_c$ the particle density is $\sim (Nm\omega^2/T_c)$ and the de
Broglie wavelength of particles
$\Lambda\sim\sqrt{\hbar^2/mT_c}$ becomes comparable with
the mean interparticle separation $\sim (Nm\omega^2/T_c)^{-1/2}$.
The result of Eq.(\ref{N02D}) is similar to that in the 3D case and it was 
first obtained by Bagnato and Kleppner \cite{Kleppner}.

\begin{figure}
\centerline{\includegraphics[width=.7\textwidth]{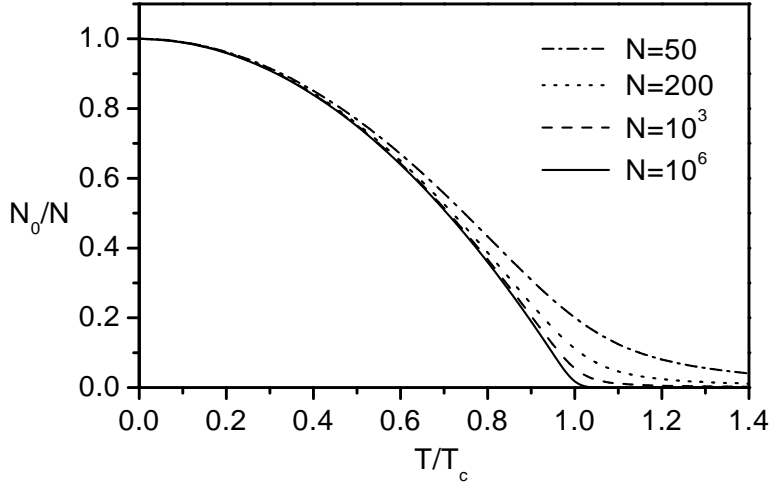}}
  \caption{The ground state population in a 2D trap versus temperature,
calculated from Eq.~(\ref{GroundPopeq}).} 
\label{Fig2D}
\end{figure}

Above $T_c$ one can omit the first term on the rhs of Eq.(\ref{GroundPopeq}).
The width of the cross-over region, {\it i.e.} the temperature interval where
both terms on the rhs of Eq.(\ref{GroundPopeq}) are equally important,  is
given by 
\begin{equation}         \label{crossover}
\frac{\Delta T}{T_c}\sim\sqrt{\frac{\ln N}{N}}\quad .
\end{equation}
The Bose-condensed fraction of particles, $N_0(T)/N$, following from 
Eq.(\ref{GroundPopeq}), is presented in Fig.~\ref{Fig2D} for various values
of $N$. For a large $N$ the cross-over region is very
narrow and one can speak of an ordinary BEC transition in an
ideal harmonically trapped 2D gas.

For the 1D Bose gas in a harmonic potential $V(z)=m\omega^2z^2/2$, the particle energy  
is $E=\hbar\omega j$, with $j$
being a non-negative integer, and the density of states is
$\rho(E)=1/\hbar\omega$. Here the integral representation of Eq.(\ref{Nint}) 
fails as the integral diverges for $\mu\rightarrow 0$. Therefore, we
should correctly take into account the discrete structure of the lowest energy
levels. In the limit $\{-\mu,\,\hbar\omega\}\ll T$ we rewrite
Eq.(\ref{Nnu}) in the form
\begin{equation}           \label{Nnu1D}
N=N_0+\frac{T}{\hbar\omega}\sum_{j=1}^{M}\frac{1}{j-\mu/\hbar\omega}+
\sum_{j=M+1}^{\infty}\frac{1}{\exp(\hbar\omega j/T-\mu/T)-1} \quad ,
\end{equation}
where the number $M$ satisfies the inequalities $1\ll M\ll
T/\hbar\omega$. The first sum is 
\begin{equation}\label{sum1}
\sum_{j=1}^{M}\frac{1}{j-\mu/\hbar\omega}=\psi(M+1-\mu/\hbar\omega)
-\psi(1-\mu/\hbar\omega)\approx
-\psi(1-\mu/\hbar\omega)+\ln(M-\mu/\hbar\omega),
\end{equation}
where $\psi$ is the digamma function. The second sum in
Eq.(\ref{Nnu1D}) can be transformed to an integral
\begin{eqnarray}   \nonumber
\sum_{j=M+1}^{\infty}\frac{1}{\exp(\hbar\omega j/T-\mu/T)-1}
\approx\frac{T}{\hbar\omega}\int_{\hbar\omega
M/T}^{\infty}\frac{{\rm d}x}{\exp(x-\mu/T)-1}\approx
-\frac{T}{\hbar\omega}\ln \frac{\hbar\omega
(M-\mu/\hbar\omega)}{T} \quad .
\end{eqnarray}
Finally, since in the limit of $|\mu|\ll T$ the chemical potential 
is related to the population of the ground state as $-\mu\approx T/N_0$, we reduce
Eq.(\ref{Nnu1D}) to the form
\begin{equation}         \label{Nnu1DFin}
N-\frac{T}{\hbar\omega}\ln\!\left(
\frac{T}{\hbar\omega}\right)=N_0-\frac{T}{\hbar\omega}\psi\!\left(1+\frac{T}{\hbar\omega
N_0}\right)\quad .
\end{equation}

\begin{figure}
\psfrag{c}[b][rc][0.75]{$1D$}
\centerline{\includegraphics[width=.7\textwidth]{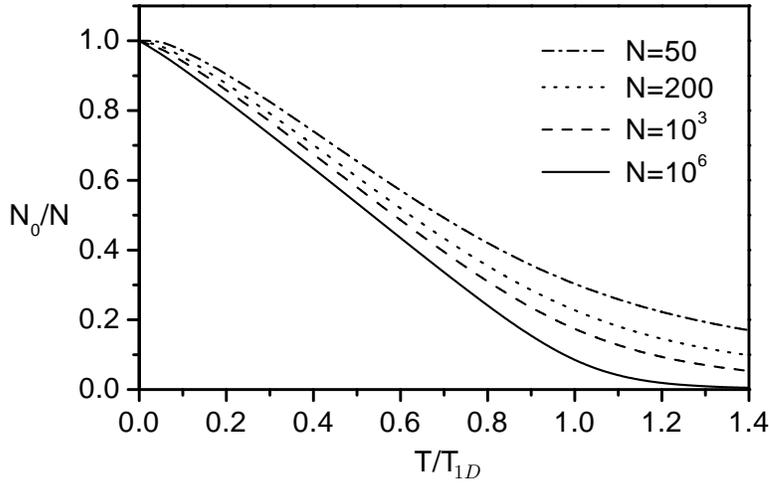}}

  \caption{\protect
The ground state population in a 1D trap versus temperature, found
from Eq.~(\ref{Nnu1DFin}).} 
\label{Fig1D}
\end{figure}

As in the 2D case, we have two regimes, with the border
between them at a temperature
\begin{equation}    \label{T1D}
T_{1D}\approx \frac{N}{\ln N}\,\hbar \omega \quad . 
\end{equation}
For temperatures below $T_{1D}$, the first term on the rhs of
Eq.(\ref{Nnu1DFin}) greatly exceeds
the second one and the ground state population behaves as
\begin{equation}  \label{N01D}
N_0\approx N-\left(\frac{T}{\hbar\omega}\right)
\ln\!\left(\frac{T}{\hbar\omega}\right) \quad . 
\end{equation}
The cross-over region is determined as the temperature interval
where both terms are equally important:
\begin{equation}   \label{crossover1D}
\frac{\Delta T}{T_{1D}}\sim\frac{1}{\ln N}.
\end{equation}
In contrast to the 3D and 2D cases, the cross-over temperature is
much lower than the degeneracy temperature $T_d\approx
N\hbar\omega$. The described results have been obtained by Ketterle and
van Druten \cite{vDK}. 
In Fig.\ref{Fig1D} we present the relative occupation of the
ground state $N_0(T)/N$ calculated from Eq.(\ref{Nnu1DFin}).

The cross-over region in the 1D case is much wider than in
2D. This is not surprising as the cross-over is
present only due to the discrete structure of the trap
levels. The quasiclassical calculation does not lead to any
sharp cross-over \cite{Kleppner}.

{\it Problem}: Describe the cross-over to the BEC regime in an ideal 2D gas of
$N$ identical bosons in a rectangular box with dimensions  $L_x,L_y$.
Find the cross-over temperature and the width of the cross-over region ({\it
E.B. Sonin, 1969}).

\section{{\it Lecture 2}. Interactions and BEC regimes in 2D trapped gases
\label{sec:IntBosegas}}

Interaction between particles drastically changes the picture of 
Bose-Einstein condensation in 2D and 1D gases. In the uniform 2D gas
a true condensate can exist only
at $T=0$ and, as mentioned in the Introduction, its absence at finite 
temperatures originates from long-wave fluctuations of the phase 
\cite{Popov}. However, as was pointed out by 
Kane and Kadanoff \cite{KK} and then proved by Berezinskii \cite{Ber}, 
there is a superfluid phase transition at sufficiently low $T$. 
Kosterlitz and Thouless \cite{KTT} found 
that this transition is associated with the formation of bound pairs of
vortices below the critical temperature 
\begin{equation}    \label{TKT}
T_{KT}=\frac{\pi\hbar^2}{2m}n_s \quad ,
\end{equation}
where $n_s$ is the superfluid density just below $T_{KT}$. This temperature is
of the order of or smaller than $T_d/4$, where $T_d=2\pi\hbar^2n/m$ is the
temperature of quantum degeneracy introduced in Lecture 1, and the notation $n$
is used in this Lecture for the 2D density. Recent Monte Carlo
calculations \cite{ProkKTT} established an exact relation between $T_{KT}$ and $T_d$
for the weakly interacting 2D Bose gas. 

Early theoretical
studies of 2D systems have been reviewed by Popov \cite{Popov} and have led to the
conclusion that below the Kosterlitz-Thouless transition temperature the Bose
liquid (gas) is characterized by the presence of a {\it quasicondensate}, that is a
condensate with fluctuating phase (see, e.g., \cite{Kagan87}). In this case the system
can be divided into blocks with a characteristic size greatly exceeding the
healing length but smaller than the phase coherence length. In each block one has
a true condensate but the phases of different blocks are not correlated with each
other.

The Kosterlitz-Thouless transition has been observed in monolayers of liquid helium
\cite{KTTexp}, and recently the observation of this transition has been reported for the
2D gas of spin-polarized atomic hydrogen on liquid helium surface \cite{Simo}.   
The Kosterlitz-Thouless transition is discussed in the lectures of Benoit
Dou\c{c}ot. 

In this Lecture we focus our attention on the interaction between particles
and on the correlation properties of 2D Bose-condensed gases. Recently, the 
2D regime was realized for trapped atomic Bose-Einstein condensates at 
MIT \cite{Ketlowd}, LENS \cite{Ing}, Innsbruck \cite{Rudi}, and JILA \cite{Eric2D}. 
This was done by a tight optical confinement of the particle motion in one direction 
\cite{Ketlowd,Ing}, or by a fast rotation of the cloud leading to an increase in the 
size of the sample in two directions \cite{Eric2D}. We will focus special attention on how the
nature of the Bose-condensed state is influenced by a finite size of the system.

\subsection{Weakly interacting regime}

We will consider weakly interacting gases with a short-range potential of
interaction between particles. In this case the total interaction energy is
equal to the sum of pair interactions and can be written as $E_{\rm
int}=N^2g/2\Omega$, where $g$ is the coupling constant for the pair
interparticle interaction, $N$ is the number of  particles, and $\Omega$ the
volume. Accordingly, the interaction energy per particle is equal to $I=ng$,
with $n$ being the 2D density. 

Let us now discuss the conditions which are required for the 2D gas
to be in the weakly interacting regime. The commonly used criterion assumes
that the mean interparticle separation ${\bar r}$ greatly exceeds the
characteristic radius of interaction between particles, $R_e$. In the 2D case 
we have ${\bar r}\approx (2\pi n)^{-1/2}$ and thus obtain the inequality
\begin{equation}    \label{crit}
nR_e^2\ll 1 \quad .
\end{equation}    

The weakly interacting regime requires that at interparticle distances of
the order of ${\bar r}$, the wavefunction of particles is not influenced by the
interaction between them. Relying on this requirement, we develop a physical
picture which will be used for finding how the criterion of the weakly
interacting regime depends on the dimensionality of the system. We consider a
box of size ${\bar r}$, which on average contains one particle. In the limit of
$T\rightarrow 0$, the particle kinetic energy is
$K\approx\hbar^2/m{\bar r}^2$. The wavefunction of the particle is not influenced by
the interparticle interaction if $K$ is much larger than the interaction
energy per particle, $I=ng$. The inequality $K\gg |I|$ immediately gives the
criterion of the weakly interacting regime in terms of the density and 
coupling constant. In the 2D case this criterion takes the form
\begin{equation}    \label{critg}
\frac{m|g|}{2\pi\hbar^2}\ll 1 \quad .
\end{equation}

In the 2D gas at $T\rightarrow 0$ the coupling constant
is (see, e.g. \cite{Popov}) 
\begin{equation}    \label{g*}
g=\frac{4\pi\hbar^2}{m}\frac{1}{\ln(1/nd_*^2)} \quad ,
\end{equation}
where a length $d_*$ depends on the shape of the interatomic potential
and in the absence of scattering resonances is of the order of $R_e$.
Then, from Eq.(\ref{critg}) we immediately arrive at the
criterion (\ref{crit}). However, on approach to a resonance, the value of $d_*$ is
quite different from $R_e$, and one should return to the criterion (\ref{critg}).        

In the dilute limit Eq.(\ref{g*}) gives $g>0$, except for the case where 
a weakly bound state of colliding atoms is present. Then the length $d_*$ is
extremely large and even at very low densities one can have the condition
$nd_*^2\gg 1$ leading to an attractive mean-field interaction ($g<0$).
However, in this case the 2D Bose gas at $T\rightarrow 0$ is unstable with
regard to collapse.    

Note that in the 3D case we have the coupling constant $g_{3D}=4\pi\hbar^2a/m$
and ${\bar r}\sim n_{3D}^{-1/3}$, with $n_{3D}$ being the 3D density, and $a$ the 3D
scattering length. The condition $K\gg |I|$ then leads to the well-known
criterion $n|a|^3\ll 1$, required for the weakly interacting regime in the
3D gas. In the absence of resonances, the 3D scattering length is of the order
of $R_e$ and this criterion is equivalent to the inequality $R_e\ll {\bar r}$. 
  
In ongoing experiments, two-dimensional atomic gases are obtained by (tightly) 
confining the motion of particles in one direction to zero point oscillations
\cite{Ketlowd,Ing,Rudi} and, in this respect, can be called quasi2D \cite{Petrov2D}. 
Kinematically the
gas is two-dimensional, but the value of an effective 2D coupling constant $g$ 
for the interparticle interaction depends on the particle motion in the tightly 
confined direction.

Let us first make a qualitative analysis of the interactions in such quasi2D gas. 
It can be viewed as a 3D gas which is uniform in two directions ($x$ and $y$), 
and is confined to zero point oscillations by a harmonic potential 
$m\omega_0^2z^2/2$ in the third direction ($z$). The quasi2D regime requires the 
inequality 
\begin{equation}      \label{quasi2D}
\hbar\omega_0\gg n|g| \quad , 
\end{equation}
where $n$ is the number of particles per unit area in the $x-y$ plane and 
represents the 2D density of the quasi2D gas. Then the distribution of the density 
in the $z$ direction is not influenced by the interactions and is given by 
\begin{equation}      \label{distr3D}
n_{3D}(z)=\frac{n}{\sqrt{\pi l_0^2}}\exp\!\left(-\frac{z^2}{l_0^2}\right)\quad ,
\end{equation} 
with $l_0=(\hbar/m\omega_0)^{1/2}$ being the harmonic oscillator length. 
The average interaction energy per particle is obtained by averaging the 3D 
interaction over the density profile in the $z$ direction:
\begin{equation}         \label{gint}
I=\frac{g_{3D}\int_{-\infty}^{\infty}n_{3D}^2(z)dz}
{\int_{-\infty}^{\infty}n_{3D}(z)dz}=ng \quad .
\end{equation}  
Eqs.~(\ref{distr3D}) and (\ref{gint}) lead to the following relation for the
effective 
coupling constant:
\begin{equation}      \label{geff}
g=\frac{2\sqrt{2}\hbar^2}{m}\frac{a}{l_0} \quad .
\end{equation}
Accordingly, the criterion of the weakly interacting regime given by Eq.(\ref{critg})
takes the form
\begin{equation}      \label{quasicrit}
l_0\gg |a| \quad .
\end{equation}
One can easily check that under conditions (\ref{quasi2D}) and (\ref{quasicrit})
the gas satisfies the 3D criterion of weak interactions, $n_{3D}|a|^3\ll 1$.

The criterion (\ref{quasicrit}) is independent of the gas density and in this 
respect is different from the criterion of the weakly interacting regime
in the purely 2D case, following from Eqs.~(\ref{critg}) and (\ref{g*}). The density
enters the problem only through the condition of the quasi2D regime given by 
Eq.(\ref{quasi2D}). However, the above analysis does not take into account the 
2D character of the particle motion at large distances between them. In the next 
section we discuss the quasi2D scattering problem and derive a more exact 
criterion of weak interactions in the quasi2D case.

\subsection{Quasi2D scattering problem}

Collisional properties of cold atoms strongly confined in one direction
have been of great interest in the studies of spin-polarized atomic hydrogen.
The interest was related to inelastic and elastic collisions in the 2D gas 
of hydrogen atoms adsorbed on liquid helium surface (see \cite{walraven} 
for review). The creation of atomic gases in the quasi2D regime 
\cite{Ketlowd,Ing,Rudi,Eric2D} opens new handles on studying 2D features of interparticle 
collisions, and provides spectacular evidence for the relation between the 
quasi2D scattering parameters and those in 3D.   
Theoretical studies of elastic and inelastic interactions  
in the quasi2D regime have been performed in \cite{Petrov2D,Petrov2Dcol,Pricoupenko},
and it has been shown how a decrease of the tight confinement transforms this 
regime into the 3D one \cite{Petrov2Dcol}.

Here we consider elastic scattering of two atoms in the quasi2D regime 
\cite{Petrov2Dcol}.
So, the atoms are (tightly) confined to zero point oscillations in the axial
($z$)  direction, and their motion in two other ($x,y$) directions is free at
a large separation in the $x-y$ plane. For a harmonic axial confinement, the
motion of  two atoms interacting with  each other via the potential $U(r)$ can
be still  separated into their relative and  center-of-mass motion. The latter
drops out of  the scattering problem. The relative motion is governed by the
potential $U(r)$  and by the potential $V_H(z)=m\omega_0^2z^2/4$ originating
from the axial  confinement with frequency $\omega_0$. In the quasi2D regime
one has the condition 
\begin{equation}     \label{quasisc}
\hbar\omega_o\gg \varepsilon \quad ,
\end{equation} 
where $\varepsilon=\hbar^2q^2/m$ and ${\bf q}$ are the energy and wavevector
of the motion in the $x-y$ plane. Therefore, the atoms are in the ground state of 
the potential $V_H(z)$ both in the incident and the scattered wave. The
wavefunction of  the relative motion satisfies the Schr\"odinger equation   
\begin{equation}     \label{Schr}  
\left(-\frac{\hbar^2}{m}\Delta+U(r)+V_H(z)-\frac{\hbar\omega_0}{2}\right)
\psi({\bf r})=\varepsilon\psi({\bf r}) \quad .   
\end{equation}   

We will consider the ultracold limit where the characteristic de Broglie 
wavelength of atoms greatly exceeds the radius of interatomic interaction $R_e$.
For the motion in the $x-y$ plane the de Broglie wavelength is $\sim 1/q$,
and we immediately obtain the inequality $qR_e\ll 1$. 
Under this condition the scattering is determined by the contribution 
of the $s$-wave for the motion in the $x-y$ plane. In the axial direction, the
atoms are tightly confined and the axial harmonic oscillator length 
$l_0$ plays the role of their axial de Broglie wavelength. Therefore, we also require
the condition 
\begin{equation}   \label{ultra}
l_0\gg R_e \quad ,
\end{equation}
which will allow us to consider only the $s$-wave for the three-dimensional relative 
motion of the atoms when they approach each other to short distances. The
condition of the quasi2D regime (\ref{quasisc}) can be also written as
$ql_0\ll 1$, and one clearly sees that Eqs.~(\ref{quasisc}) and (\ref{ultra}) 
automatically lead to the inequality $qR_e\ll 1$. 
  
The scattering amplitude is defined through the asymptotic form of the
wavefunction $\psi$ at an infinite separation $\rho$ in the $x-y$ plane,
where it is represented as a superposition of an incident and scattered wave:
\begin{equation} \label{Asym}  
\psi({\bf r})\approx\varphi_0(z)\left(e^{i\bf{q.}\mbox{\footnotesize\boldmath$\rho$}}-f(q,\phi)
\sqrt{\frac{i}{8\pi q\rho}}e^{iq\rho}\right) \quad , 
\end{equation} 
where $\varphi_0(z)=(1/2\pi l_0^2)^{1/4}\exp(-z^2/4l_0^2)$ is the eigenfunction
of the ground state in the potential $V_H(z)$, and $\phi$ is the scattering
angle. The $s$-wave scattering is circularly symmetric in the $x-y$ plane, and
the scattering amplitude is independent of $\phi$. Note that $f(q,\phi)$ in
Eq.(\ref{Asym}) is different by a factor of $-\sqrt{8\pi q}$ from the
definition of the 2D scattering amplitude used in \cite{LLQ}.   

Relying on the condition (\ref{ultra}) we will express
the scattering amplitude through the 3D scattering length.
At interparticle distances $r\gg R_e$ the relative motion in the $x-y$ plane
is free, and the motion along the $z$ axis is governed only by the harmonic
potential $V_H(z)$. Then, the solution of Eq.(\ref{Schr}) with
$U(r)=0$ can be expressed through the Green function $G_{\varepsilon}
({\bf r},{\bf r}')$ of this equation. Retaining only the $s$-wave for 
the motion in the $x-y$ plane, we have 
\begin{equation}    \label{Green}  
\psi({\bf r})=\varphi_0(z)J_0(q\rho)-fG_{\varepsilon}({\bf r},0)/\phi_0(0)
\quad .  
\end{equation}  
For $\rho\rightarrow\infty$ the Green function is $G({\bf r},0)=
\phi_0(z)\phi_0(0)\sqrt{(i/8\pi q\rho)}\exp{(iq\rho)}$ and Eq.(\ref{Green})
gives the $s$-wave of Eq.(\ref{Asym}).

The condition $l_0\gg R_e$ ensures that the relative motion of atoms in the
region of interatomic interaction is not influenced by the axial (tight)
confinement. Therefore, the wavefunction $\psi({\bf r})$ at distances 
$R_e\ll r\ll l_0$ differs only by a normalization coefficient from the 3D 
wavefunction of free
motion at zero energy, $\psi_{3D}(r)$. Writing this coefficient as
$\eta\varphi_0(0)$ and recalling that for $r\gg R_e$ one has 
$\psi_{3D}(r)=(1-a/r)$, we obtain 
\begin{equation}   \label{3D}  
\psi(r)\approx\eta\varphi_0(0)\left(1-\frac{a}{r}\right) \quad ; \qquad R_e\ll r\ll l_0 \quad .   
\end{equation} 
Eq.(\ref{3D}) contains the 3D scattering length $a$ and 
serves as a boundary condition for $\psi({\bf r})$ (\ref{Green})
at $r\rightarrow 0$. 

For $r\rightarrow 0$, a straightforward calculation of the Green function 
$G({\bf r},0)$ yields \cite{Petrov2D,Petrov2Dcol}
\begin{equation} \label{lim}
G_{\varepsilon}(r,0)\approx\frac{1}{4\pi
r}+\frac{1}{2(2\pi)^{3/2}l_0}\left[\ln\!\left(\frac{B\hbar\omega_0}{\pi\varepsilon}\right)+i\pi\right]
\quad ,
\end{equation}
where $B\approx 0.915$.
With the Green function (\ref{lim}), the wavefunction (\ref{Green}) at 
$r\rightarrow 0$ should coincide with $\psi(r)$ (\ref{3D}). This gives the 
coefficient $\eta$ and the scattering amplitude: 
\begin{equation}    \label{Amplfin} 
f(\varepsilon)=4\pi\varphi_0^2(0)a\eta=\frac{2\sqrt{2\pi}}
{l_0/a+(1/\sqrt{2\pi})[\ln(B\hbar\omega_0/\pi\varepsilon)+i\pi]} \quad .   
\end{equation}
One can see from Eq.(\ref{Amplfin}) that the scattering amplitude is a 
universal function of the parameters $a/l_0$ and $\varepsilon/\hbar\omega_0$.

The 2D kinematics of the relative motion at interatomic distances $\rho>l_0$ 
manifests itself in the appearance of logarithmic dependence of the scattering
amplitude on  $\varepsilon/2\hbar\omega_0$. The quasi2D scattering amplitude 
can be represented in the purely 2D form obtained, for example, in \cite{LLQ}:
\begin{equation}    \label{Purf} 
f(q)=\frac{2\pi}{\ln(1/qd_*)+i\pi/2} \quad . 
\end{equation} 
In the purely 2D case a characteristic length $d_*$ depends on the shape of the
interatomic potential $U(r)$. For the considered quasi2D regime this length is 
expressed through $l_0$ and the 3D scattering length $a$:
\begin{equation}   \label{dquasi}
d_{*}=\frac{d}{2}\,e^C = \sqrt{\frac{\pi}{B}}\,l_0\exp\!\left(-\sqrt{\frac{\pi}{2}}\,\frac{l_0}{a}\right) \quad,
\end{equation}
where $C$ is the Euler's constant.

We thus conclude that the scattering problem in the quasi2D regime is 
equivalent to the scattering in an effective purely 2D potential which leads
to the same value of $d_*$. 
For a negative $a$ satisfying the condition $|a|\ll l_0$, the effective
potential is a shallow well which has a depth $|V_0|$ and a radius $l_0$. This
shallow well supports a weakly bound state with an exponentially small binding
energy $\varepsilon_0$, which leads to an exponentially large $d_*$ 
as follows from Eq.(\ref{dquasi}). As a result, we have a resonance energy 
dependence of the scattering amplitude $f$
at a fixed ratio $a/l_0$, and a resonance behavior of $f$ as a function
of $a/l_0$ at a fixed $\varepsilon/\hbar\omega_0$. 

Equation (\ref{Amplfin}) shows that the resonance is achieved at 
$a=a_*(\varepsilon)=-\sqrt{2\pi}l_0/\ln{(B\hbar\omega_0/\pi\varepsilon)}$.
One can think of observing the resonance dependence of $f$ on  $a/l_0$  
by measuring the rate of elastic collisions, which is proportional to 
$|f|^2$. For example, one can keep temperature and
$\omega_0$ constant and vary $a$ by using Feshbach resonances. This 
will be a striking difference from the 3D case, where the rate constant 
of elastic collisions monotonically increases with $a^2$. 

Stanford \cite{Chu1,Chu2} and ENS \cite{Christ1,Christ2} experiments with 
a thermal gas of Cs atoms
tightly confined in one direction, observed a pronounced deviation of 
collisional rates from the 3D behavior. In these experiments the temperature
was of the order of $\hbar\omega_0$ and, in this respect, they were in 
between the quasi2D and 3D regimes.  

The scattering amplitude $f$ determines the coupling constant $g$ for a fixed collision
energy $\varepsilon$. Away from the resonance, one may
omit the imaginary part in Eq.(\ref{Amplfin}) and $f$ becomes real. Using 
Eq.(\ref{Schr}) one can establish that in the ultracold limit the interaction 
energy for a pair of atoms is equal to $g/\Omega$, where the coupling constant 
is given by
\begin{equation}       \label{coupling}
g=\int\!\!d^3r\,\varphi_0(z)U(r)\psi({\bf r})\quad,
\end{equation}
and $\Omega$ is the volume of the system. Accordingly, the total interaction energy
is equal to $(g/\Omega)$ multiplied by the number of pairs $N^2/2$, as stated in
the beginning of this Lecture.

The main contribution to the integral in Eq.(\ref{coupling}) comes from interparticle
distances $r\sim R_e$. Therefore, one may put $\varphi_0(z)\approx\varphi_0(0)$ and
$\psi(r)\approx \eta\varphi_0(0)\psi_{3D}(r)$. Then, using the well-known result
$\int\!\!d^3r\,\psi_{3D}(r)U(r)=4\pi\hbar^2a/m$ and Eq.(\ref{Amplfin}), we obtain
\begin{equation}     \label{gquasi2D}
g=\frac{\hbar^2 f(\varepsilon)}{m}=\frac{4\pi\hbar^2}{m}\frac{1}
{\sqrt{2\pi}l_0/a+\ln(B\hbar\omega_0/\pi\varepsilon)} \quad . 
\end{equation}
As one can see, the coupling constant for the interaction between particles depends
on their relative energy. In a thermal gas, averaging the coupling constant over
the energy distribution of particles simply leads to the replacement of $\varepsilon$
by $T$ in Eq.(\ref{gquasi2D}).

In a Bose-condensed 2D gas, similarly to the 3D case (see, e.g. \cite{AGD}), 
to zero order in perturbation theory the coupling constant is equal to
the vertex of interparticle interaction in vacuum at zero momenta
and frequency ${\bar \omega}=2\mu$. For low ${\bar \omega}>0$ this vertex
coincides with the amplitude of scattering at relative energy ${\bar \omega}$
and, hence, is given  by Eq.(\ref{quasi2D}) with $\varepsilon=2\mu$. This
certainly requires the quasi2D condition $\hbar\omega_0\gg\mu$. For $\mu<0$,
analytical continuation of Eq.(\ref{gquasi2D}) to $\varepsilon<0$ leads to the
replacement  $\varepsilon\rightarrow |\varepsilon|=2|\mu|$.

We can now analyze the criterion of the weakly interacting regime for the
ultracold quasi2D gas, assuming that it is dilute and the condition 
$nR_e^2\ll 1$ is fulfilled. Since the scattering amplitude $f$ and, hence, 
the coupling constant $g$ can be written in  
the purely 2D form (see Eq.(\ref{Purf})), the weakly interacting regime 
requires the inequality $m|g|/2\pi\hbar^2\ll 1$ given above by Eq.(\ref{critg}).
For $a>0$, the quasi2D resonance is absent and this inequality is satisfied
for any ratio $a/l_0$. For a negative $a$, the system should be far away from
the resonance and one should use Eq.(\ref{gquasi2D}) to make sure that the
condition (\ref{critg}) is satisfied. This is demonstrated in Fig.3 where 
we present the parameter $mg/2\pi\hbar^2$ as a function of the ratio 
$a/l_0$. The results are obtained from 
Eq.(\ref{gquasi2D}) for a fixed ratio $\hbar\omega_0/\varepsilon=10^3$. 
In the vicinity of the resonance, where the criterion
$m|g|/2\pi\hbar^2\ll 1$ is not satisfied, they are shown by the dashed curves.   

\begin{figure}
\centerline{\includegraphics[width=.7\textwidth]{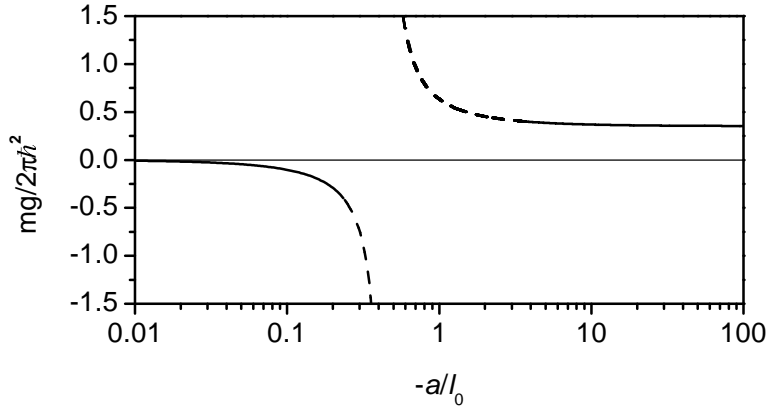}}
\caption{The parameter $mg/2\pi\hbar^2$ versus $a/l_0$, obtained from 
Eq.(\ref{gquasi2D}) (see text).} 
\label{Fig21}
\end{figure} 

\subsection{True BEC at $T=0$}

We start the discussion of BEC regimes in (quasi)2D weakly interacting gases
with the case of zero temperature. From this point on we 
consider only repulsive
interaction between particles ($g>0$). The behavior of attractively interacting
Bose gases and, in particular, the problem of collapses are beyond the scope of 
these lectures. 

In the second quantization the Hamiltonian of
the system reads: 
\begin{equation}     \label{tildeH}
\hat H=\int\!\!d{\bf r}\,\hat\Psi^{\dagger}\left(
-\frac{\hbar^2}{2m}\Delta+V({\bf r})+\frac{g}{2}\hat\Psi^{\dagger}
\hat\Psi\right)\hat\Psi \quad ,
\end{equation}
where $\hat\Psi({\bf r},t)$ is the Heisenberg field operator of the atoms,
$V({\bf r})$ is an external (trapping)
potential, and all field operators in the integrand are taken at the same
position ${\bf r}$ and for the same time $t$. The Heisenberg equation of motion
for the field operator takes the form:  
\begin{equation}       \label{Heistot}
i\hbar\frac{\partial\hat\Psi}{\partial t}=[\hat\Psi,\hat H]= 
\left( -\frac{\hbar^2}{2m}\Delta+V({\bf r})+g\hat\Psi^{\dagger}\hat\Psi
\right)\hat\Psi \quad .
\end{equation}

We now assume {\it a priori} that there is a true condensate at $T=0$, and the
condensate density $n_0$ is much larger than the density of non-condensed
particles $n'$. Accordingly, we represent the field operator as a sum of the
non-condensed part $\hat\Psi^{\prime}$ and the condensate wavefunction
$\Psi_0=\sqrt{n_0}$ which is a $c$-number:
\begin{equation}       \label{operator}
\hat\Psi=\hat\Psi^{\prime}+\Psi_0 \quad .
\end{equation}
At equilibrium, the time dependence of the condensate wavefunction is reduced
to $\Psi_0\propto\exp{(-i\mu t/\hbar)}$, where $\mu$ is the chemical potential.
Then, taking average of both sides of Eq.(\ref{Heistot}) and omitting the
contribution of non-condensed particles, we obtain the Gross-Pitaevskii
equation for the condensate wavefunction \cite{GP1,GP2}: 
\begin{equation}        \label{GP}
\left(-\frac{\hbar^2}{2m}\Delta+V({\bf r})+g|\Psi_0|^2-\mu\right)\Psi_0=0
\quad .
\end{equation} 
The function $\Psi_0$ is normalized by the condition 
\begin{equation}    \label{normo}
\int\!\!d{\bf r}\,|\Psi_0|^2=N\quad,
\end{equation}
which gives a relation between the number of particles $N$ and chemical
potential $\mu$. 

Equations (\ref{Heistot}) and (\ref{GP}) immediately lead to the equation of
motion for the non-condensed part of the field operator:      
\begin{equation}        \label{Heisnon} 
i\hbar\frac{\partial\hat\Psi^{\prime}}{\partial t}=
\left( -\frac{\hbar^2}{2m}\Delta+V({\bf
r})+2g|\Psi_0|^2\right)\hat\Psi^{\prime}
+g\Psi_0^2\hat\Psi^{{\prime}{\dagger}}\quad . \end{equation}
We then use the Bogoliubov transformation (see \cite{LLS}) generalized for the spatially
non-uniform case \cite{dG} and express $\Psi^{\prime}$ through the eigenmodes
of elementary excitations:
\begin{equation}       \label{Bogtr}
\Psi^{\prime}=\exp\!\left(-i\mu t/\hbar\right) \sum_{\nu}\left[ u_{\nu}\hat
a_{\nu}\exp\!\left(-i\varepsilon_{\nu}t/\hbar\right)-v_{\nu}^*\hat
a_{\nu}^{\dagger}\exp\!\left(i\varepsilon_{\nu}t/\hbar\right)\right] \quad .
\end{equation}
Here $\hat a_{\nu},\hat a_{\nu}^{\dagger}$ are annihilation and creation
operators of the excitations, and $\varepsilon_{\nu}$ are their eigenenergies.
Excitations are characterized by a set of quantum numbers $\nu$, and their
Bogoliubov $\{u_\nu,v_\nu\}$ functions are normalized by the condition 
\begin{equation}     \label{normex}
\int\!\!d{\bf r}\,\left(|u_{\nu}({\bf r})|^2-|v_{\nu}({\bf r})|^2\right)=1\quad .
\end{equation}
Commuting both sides of Eq.(\ref{Heisnon}) with the operators $\hat a_{\nu}$
and $\hat a_{\nu}^{\dagger}$, we arrive at the Bogoliubov-de Gennes equations
for the excitation energies and wavefunctions. Assuming that aside from the
factor $\exp(-i\mu t/\hbar)$, the condensate wavefunction 
$\Psi_0$ is real, these equations read:
\begin{eqnarray} 
\left(-\frac{\hbar^2}{2m}\Delta+V({\bf
r})+g|\Psi_0|^2-\mu\right)f^+_{\nu} & = & \varepsilon_{\nu}f^-_{\nu}  
\label{BdG+}  \\
\left(-\frac{\hbar^2}{2m}\Delta+V({\bf
r})+3g|\Psi_0|^2-\mu\right)f^-_{\nu} & = & \varepsilon_{\nu}f^+_{\nu} \quad ,  
\label{BdG-}
\end{eqnarray}
where the functions $f_{\nu}^{\pm}=u_{\nu}\pm v_{\nu}$.

In the spatially uniform case Eq.(\ref{GP}) is reduced to the expression for
the chemical potential through the condensate density: $\mu=n_0g$. Excitations
are characterized by their wave vector ${\bf k}$, and Eqs.~(\ref{BdG+}), 
(\ref{BdG-}) give the eigenfunctions
\begin{equation}     \label{fpmun}
f_k^{\pm}=u_k\pm
v_k=\frac{1}{\Omega}\left(\frac{\varepsilon_k}{E_k}\right)^{\pm 1/2}
\exp{(i{\bf k.r})} 
\end{equation}
and lead to the Bogoliubov energy spectrum
\begin{equation}     \label{Bogsp}
\varepsilon_k=\sqrt{E_k^2+2\mu E_k} \quad ,
\end{equation}
where $E_k=\hbar^2k^2/2m$ is the energy of a free particle. For small momenta,
excitations are phonons characterized by a linear dispersion law
$\varepsilon_k=\hbar c_sk$, where $c_s=\sqrt{\mu/m}$ is the velocity of sound.
High-momentum excitations are single particles with $\varepsilon_k=E_k+\mu$.
Characteristic momenta at which the linear spectrum transforms into the
quadratic one are of the order of $1/l_c$, where $l_c=\hbar/\sqrt{m\mu}$ is
called correlation or healing length. The corresponding excitation energies
are $\sim \mu$. Note that the criterion (\ref{critg}) of the weakly interacting
regime is equivalent to the condition $l_c\gg {\bar r}$. In other words, 
there are many particles under the healing length.

We now calculate the density of non-condensed particles $n'$ and the
single-particle correlation function. Using Eqs.~(\ref{Bogtr}) and
(\ref{fpmun}) we obtain
\begin{equation}        \label{nprim}
n'=\langle\hat\Psi^{{\prime}{\dagger}}({\bf r})\hat\Psi^{\prime}({\bf r})\rangle=
\int\!\!\frac{\Omega d^2 k}{(2\pi)^2}\,v_k^2=\frac{mg}{4\pi\hbar^2}n_0 \quad .
\end{equation}
One clearly sees that the ratio $n'/n_0$ is small. It is proportional to the
small parameter  $mg/2\pi\hbar^2$ required for the weakly interacting regime
in the 2D Bose gas.

The single-particle correlation function is $g_1({\bf r}',{\bf
r}'')=\langle\hat\Psi^{\dagger}({\bf r}')\hat\Psi({\bf r}'')\rangle$ and it depends only on
the relative coordinate $r=|{\bf r}'-{\bf r}''|$. On the basis of
Eqs.~(\ref{operator}), (\ref{Bogtr}), and (\ref{fpmun}) we obtain
\begin{equation}      \label{g12D}
g_1(r)=n_0+\int\!\!\frac{\Omega d^2k}{(2\pi)^2}\,v_k^2\exp(i{\bf k.r})
=n_0\left[ 1+\frac{mg}{2\pi\hbar^2}I_1\!\left(r \over l_c\right)K_1\!\left(r \over l_c\right)\right]\quad ,
\end{equation}
where $I_1$ and $K_1$ are the growing and decaying modified Bessel functions,
respectively. The result of Eq.(\ref{g12D}) is displayed in Fig.4. 
For $r\ll l_c$ the single-particle correlation function is close to the total
density $n$. At distances $r\gg l_c$, Eq.(\ref{g12D}) gives
\begin{equation}    \label{g12Dass}
g_1(r)=n_0\left( 1+\frac{mg}{4\pi\hbar^2}\frac{l_c}{r}\right)\quad ;\qquad r\gg
l_c \quad . 
\end{equation}
For $r\rightarrow\infty$ the second term in the brackets vanishes and $g_1$
is tending to $n_0$. Thus, there is a long-range order in the system.
This justifies our initial assumption of the presence of a true BEC in the 2D
weakly interacting gas at $T=0$. Note that $g_1$ drops from $n$ to $n_0$ at
distances of the order of the healing length $l_c$.

In order to go beyond the Bogoliubov approach and find, for example, 
corrections to the excitation energies, which are small as $mg/2\pi\hbar^2$,
one may proceed along the lines of the Beliaev theory \cite{Beliaev} developed
for 3D Bose-condensed gases at $T=0$.

\begin{figure}
\centerline{\includegraphics[width=.7\textwidth]{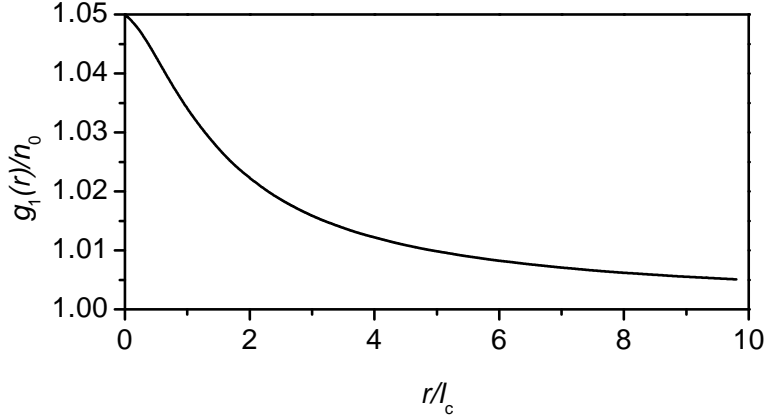}}
\caption{Normalized single-particle correlation function $g_1/n_0$ versus $r/l_c$,
for a uniform 2D Bose gas at zero temperature.} 
\label{Fig22}
\end{figure}

\subsection{Quasicondensate at finite temperatures}

As we already mentioned,
the finite-temperature uniform 2D Bose gas is characterized by the absence of
a true Bose-Einstein condensate and long-range order. 
To gain insight into the nature of this
phenomenon we first try to calculate the density of non-condensed particles,
assuming that there is a true BEC and one can use the Bogoliubov approach
described in the previous subsection. Then, at finite temperatures, from 
Eqs.~(\ref{Bogtr}) and (\ref{fpmun}) we obtain
\begin{equation}        \label{nprimsum}
n^{\prime}=\langle\Psi^{{\prime}{\dagger}}({\bf r})\Psi^{\prime}({\bf r})\rangle=
n^{\prime}_v+n^{\prime}_T \quad .
\end{equation}
The vacuum (temperature-independent) contribution $n^{\prime}_v$ is given
by Eq.(\ref{nprim}), and the thermal contribution is determined by the relation
\begin{equation}       \label{nprimT}
n^{\prime}_T=\int\!\!\frac{\Omega d^2k}{(2\pi )^2}\,(v_k^2+u_k^2)N_k=
\int\!\!\frac{d^2k}{(2\pi)^2}\,\left(\frac{E_k+\mu}{\varepsilon_k}\right)N_k \quad ,
\end{equation}
where $N_k=[\exp(\varepsilon_k/T)-1]^{-1}$ are equilibrium occupation
numbers for the excitations. For $k\rightarrow 0$ the occupation numbers are
$N_k\approx T/\varepsilon_k$, and the excitation energy is
$\varepsilon_k\propto k$. Therefore, the integrand in the rhs of
Eq.(\ref{nprimT}) behaves as $d^2k/k^2$ and the integral is divergent at small
momenta, which rules out the assumption of the presence of a true condensate.
The origin of this infrared divergence is related to long-wave fluctuations
of the phase of the condensate wavefunction, which can be understood turning to the
density-phase representation for the field operators:     
\begin{equation}\label{overview:dens-phase}
\hat{\Psi}=\exp
(i\hat{\phi})\sqrt{\hat{n}},\quad \hat{\Psi}^\dagger
=\sqrt{\hat{n}}\exp (-i\hat{\phi})\quad .
\end{equation}
Here the density and phase operators are real and satisfy
the commutation relation
\begin{equation}\label{overview:comrelphase}
[\hat{n}({\bf r}),\hat{\phi}({\bf r}^\prime)]=i\delta({\bf
r}-{\bf r}^\prime) \quad .
\end{equation}

We now present a general approach which is based on the Hamiltonian
(\ref{tildeH})  and includes the presence of an external trapping potential
$V({\bf r})$. We assume {\it a priori} and justify later that fluctuations of
the density are small. Substituting Eqs.~(\ref{overview:dens-phase}) into
Eq.~(\ref{Heistot}) and separating real and imaginary parts, we get the coupled 
continuity and Euler hydrodynamic equations for the density and velocity
$\hat{\bf v}=(\hbar/m)\nabla \hat{\phi}$:
\begin{eqnarray}
-\hbar\frac{\partial \sqrt{\hat{n}}}{\partial
t}&=&\frac{\hbar^2}{2m}(\Delta\hat\phi\sqrt{\hat n}+
2\nabla\hat\phi\nabla\sqrt{\hat{n}})\quad ,\label{overview:Continuity}\\
-\hbar \frac{\partial \hat{\phi}}{\partial t}\sqrt{\hat n}&=&\frac{\hbar^2}{2m}
(\nabla\hat{\phi})^2\sqrt{\hat
n}-\frac{\hbar^2}{2m}\Delta\sqrt{\hat n}+V({\bf
r})\sqrt{\hat n}+g\hat{n}^{3/2} \quad . \label{overview:Euler} 
\end{eqnarray}

For small density fluctuations, Eq.(\ref{overview:Continuity}) shows that
fluctuations 
of the phase gradient are also small. Writing the density operator as 
$\hat{n}({\bf r})=n({\bf r})+\delta\hat{n}
({\bf r})$ and shifting the phase by $-\mu t/\hbar$, we then
linearize Eqs.~(\ref{overview:Continuity}-\ref{overview:Euler}) with
respect to $\delta\hat{n}$, $\nabla\hat{\phi}$ around the
stationary solution $\hat{n}=n$, $\nabla\hat{\phi}=0$.
The zero order terms give the Gross-Pitaevskii equation for the mean density
$n$: 
\begin{equation}\label{overview:GPE}
-\frac{\hbar^2}{2m}\frac{\nabla^2\sqrt{{n}}}{\sqrt{{n}}}+V({\bf
r})+gn=\mu \quad ,
\end{equation}
and the first order terms provide equations for the density and
phase fluctuations:
\begin{eqnarray}
\hbar\,\partial(\delta\hat{n}/\sqrt{n})/\partial
t&=&(-\hbar^2\nabla^2/2m+V({\bf
r})+gn-\mu)(2\sqrt{n}\,\hat{\phi})\quad ,\label{overview:Continuitylin}\\
-\hbar\,\partial(2\sqrt{n}\,\hat{\phi})/\partial t&
=&(-\hbar^2\nabla^2/2m+V({\bf
r})+3gn-\mu)(\delta\hat{n}/\sqrt{n})\quad .\label{overview:Eulerlin}
\end{eqnarray}

Solutions of Eqs.~(\ref{overview:Continuitylin}-\ref{overview:Eulerlin})
for $\delta\hat{n}$, $\nabla\hat\phi$ are obtained in terms of elementary
excitations: 
\begin{eqnarray}
\delta\hat{n} ({\bf r})&=&n({\bf r})^{1/2}\sum_\nu
f_{\nu}^-({\bf r})e^{-i\varepsilon_\nu
t/\hbar}\,\hat{a}_\nu+{\rm h.c.}\quad ,\label{overview:expansiondens}\\
\hat{\phi}({\bf r})&=&[4n({\bf r})]^{-1/2}\sum_\nu
-if_{\nu}^+({\bf r})e^{-i\varepsilon_\nu
t/\hbar}\,\hat{a}_\nu+{\rm
h.c.}\quad ,\label{overview:expansionphase}
\end{eqnarray}
where the eigenfunctions $f_{\nu}^{\pm}$ obey the Bogoliubov-de Gennes 
equations (\ref{BdG+}) and (\ref{BdG-}), with $|\Psi_0|^2$ replaced by 
$n$. In the previous subsection these functions were introduced as
eigenfunctions of elementary excitations of a true Bose-Einstein 
condensate. We thus see that the assumption of small density 
fluctuations is sufficient for having the Bogoliubov wavefunctions and 
spectrum of the excitations, irrespective of the presence or absence of 
a true condensate. Note that this statement holds for both uniform and trapped
Bose gases in the weakly interacting regime in any dimension.

In the uniform case the chemical potential is $\mu=gn$,
and the excitation wavefunctions and spectrum are given by 
Eqs.~(\ref{fpmun}) and (\ref{Bogsp}). As discussed above, the characteristic
excitation energies at which the character of the spectrum transforms from the
phonon into the single-particle one are of the order of $\mu$.
We, therefore, separate the energy space into two regions: low-energy
(phonon) part with $\varepsilon_k<\mu$, and high-energy (free-particle) part
where  $\varepsilon_k>\mu$. The operators of the density and phase
fluctuations are then represented as:
\begin{eqnarray}   
\delta \hat{n} &=& \delta \hat{n}_p+\delta \hat{n}_f \quad , \label{nop}  \\ 
\hat\phi &=& \hat\phi_p+\hat\phi_f \quad ,  \label{phiop}
\end{eqnarray}
where the indices $p$ and $f$ stand for the phonon ($\varepsilon<\mu$)
and free-particle ($\varepsilon>\mu$) parts, respectively. 

Fluctuations originating from the high-energy part are small. This is seen 
from the calculation of the density of this part of the gas,
$<\hat\Psi_f^{\dagger}\hat\Psi_f>$, where the operator $\hat\Psi_f$ accounts
for both the density and phase fluctuations and is given by the Bogoliubov
transformation (\ref{Bogtr}) in which the summation is performed only over
excitations with energies $\varepsilon_k>\mu$. At energies significantly larger 
than $\mu$ the function $v_k=(f_k^+ -f_k^-)/2 \rightarrow 0$, and
$u_k=(f_k^++f_k^-)/2 \rightarrow \Omega^{-1/2}\exp (i{\bf k}\cdot{\bf r})$. The
excitation energy is $\varepsilon_k\approx E(k)+\mu$ and, hence, the operator
$\hat\Psi_f$ describes an ideal thermal gas of Bose particles with chemical potential 
equal to $-\mu$. The density of this gas is exponentially small at $T<\mu$. 
In the 2D case, for higher temperatures it is equal to 
\[
\langle\hat\Psi_f^{\dagger}\hat\Psi_f\rangle\approx
\int_{\varepsilon_k>\mu}\!\!\frac{d^2k}{(2\pi )^2}\,N_k \,
< \, n\frac{T}{T_d}\ln\left(\frac{T}{\mu}\right)\quad ,
\]
and is much smaller than $n$ assuming that $T$ is well below 
the temperature of quantum degeneracy $T_d=2\pi\hbar^2n/m$.

The low-energy fluctuations of the density at $T\ll T_d$ are also small. This
follows from the calculation of the density-density correlation function
$\langle \delta\hat{n}_{s}({\bf r})\delta\hat{n}_{s}(0)\rangle$. For the 2D gas,
a straightforward calculation using Eq.(\ref{overview:expansiondens}) yields
\begin{equation}\label{overview:densitycorr}
\frac{\langle\delta\hat n_p({\bf r})\delta\hat
n_p(0)\rangle}{n^2}=\frac{1}{n\Omega}\sum_{\varepsilon_k<
\mu}\frac{E(k)}{\varepsilon (k)}\left(2N_k+1\right) \cos({\bf k.r})\,
<\, {\rm max}\left\{\frac{T}{T_d},\frac{mg}{4\pi\hbar^2}
\right\}\ll 1 \quad .
\end{equation}
Similarly, one finds that the density-phase correlation functions 
$\langle\frac{\delta\hat n_p({\bf r})}{n}\hat\phi_p(0) \rangle$ and
$\langle\hat\phi_p({\bf r})\frac{\delta\hat n_p(0)}{n}\rangle$ are small for $T\ll T_d$.

To zero order in perturbation theory, omitting small high-energy fluctuations
and small low-energy fluctuations of the density, the single-particle
correlation function is found by using the field operator in the form:
\begin{equation}     \label{overview:fieldexpansion}
\hat\Psi=\sqrt{n}\exp(i\hat\phi_p) \quad .
\end{equation}
Relying on the Taylor expansion of the exponent one proves directly that 
$\langle \exp\{i\phi_p\} \rangle=\exp(-\langle\phi_p^2\rangle/2)$. Then, for the single-particle
correlation function we obtain:
\begin{equation}      \label{g1gen0}
g_1(r)=\langle\hat\Psi^{\dagger}({\bf r})\hat\Psi(0)\rangle=n\exp\left(-\frac{1}{2}
\langle(\hat\phi_p(0)-\hat\phi_p({\bf r}))^2\rangle\right) \quad ; \qquad T\ll T_d \quad .
\end{equation}

In the 2D case, vacuum low-energy (long-wave) fluctuations of the phase are
small as $mg/2\pi\hbar^2$. However, thermal phase fluctuations are large for
$r\rightarrow\infty$. On the basis of Eq.(\ref{overview:expansionphase}), 
we find the following asymptotic expression for the mean square phase
fluctuations:   
\begin{equation}          \label{overview:phasecorr2D}
\left\langle\left[\hat{\phi}_p({\bf
r})-\hat{\phi}_p({\bf
0})\right]^2\right\rangle_T\approx \frac{2T}{T_d} \ln\!\left(
\frac{r}{\lambda_T}\right)\quad ;\qquad r\gg\lambda_T \quad . 
\end{equation}  
where $\lambda_T$ is equal to the healing length $l_c$ at $T\gg\mu$, and
to the thermal de Broglie wavelength of phonons $\hbar c_s/T$ for $T\ll\mu$. 
Accordingly, the correlation function $g_1(r)$ undergoes a slow power law decay
at large distances: 
\begin{equation}       \label{powerlaw}
g_1(r)=n\left(\frac{\lambda_T}{r}\right)^{T/T_d}\quad ; \qquad r\gg\lambda_T \quad .
\end{equation}
This is drastically different from the situation in 3D Bose-condensed gases,
where fluctuations are small at any distance and $g_1$ is tending to
the condensate  density as $r\rightarrow\infty$. We thus see that just
long-wave thermal fluctuations of the phase destroy the long-range order and
true BEC in finite-temperature 2D Bose gases. 

The low-temperature behavior of the single-particle correlation function,
showing its power law decay at large $r$, was first obtained by Kane and
Kadanoff \cite{KK}. The idea of dividing the system into slow (low-energy) and
fast (high-energy) parts belongs to Popov who developed a perturbation theory
for Bose systems on these grounds \cite{Popov}. In the described hydrodynamic
approach, for obtaining perturbative corrections to Eq.(\ref{g1gen0}) one
should expand the continuity and Euler equations (\ref{overview:Continuity}) 
and (\ref{overview:Euler}) up to second order in the density and phase 
fluctuations, which provides a correction to the stationary solution. One
should then include the low-energy density fluctuations and
the high-energy fluctuations in the expression for the
field operator. An expression for $g_1(r)$ in the uniform 2D case, which contains
these corrections and is obtained on the basis of the Popov theory, is given
in \cite{Kagan00}.

The distance at which the mean square phase fluctuations become of the order
of unity and the single-particle correlation function significantly decreases,
is called the phase coherence length $l_{\phi}$. 
From Eq.(\ref{overview:phasecorr2D}) we obtain 
\begin{equation}        \label{lphi2D}
l_{\phi}=\lambda_T\exp\left(\frac{T_d}{2T}\right)
\end{equation}
and clearly see that at $T\ll T_d$ the phase coherence length greatly exceeds
the healing length $l_c$. This means that the system can be divided into blocks
of size $\tilde L$ which is chosen such that $l_c\ll \tilde L\ll l_{\phi}$.
Then, using  Eqs.~(\ref{g1gen0}) and (\ref{overview:phasecorr2D}) we can make
sure that correlation properties inside each block are the same as in a genuine
Bose-condensed gas. We thus conclude that there is a true condensate in each
block. However, the phases of different blocks are not correlated with each
other. Therefore, this system got the term {\it quasicondensate}, or
condensate with fluctuating phase \cite{Kagan87}. For the existence of the
quasicondensate, it is crucial that the density fluctuations are small on any
distance scale.

\subsection{True and quasicondensates in 2D traps}

Bose-Einstein condensation in trapped 2D gases is qualitatively 
different from that in the uniform case. As was discussed in Lecture 1,
for an ideal 2D gas the change of the density of states due to a harmonic 
confining potential, leads to a macroscopic occupation of the ground state 
of the trap (ordinary BEC) at temperatures $T<T_c\approx\!N^{1/2}\hbar\omega$, 
where $N$ is the number of particles, and $\omega$ the trap frequency
\cite{Kleppner}. Thus, there is a question of whether an interacting 
trapped 2D gas supports the ordinary BEC or the Kosterlitz-Thouless type of 
a cross-over to the BEC regime. Related studies are now underway,
and for large 2D samples one expects the Kosterlitz-Thouless cross-over.
However, irrespective of the type of the BEC cross-over, the critical 
temperature will be always comparable with $T_c$ of an ideal gas.
On approach to $T_c$ from above, the gas density is
$n_c \sim N/R_{T_c}^2$, where $R_{T_c}\approx\sqrt{T_c/m\omega^2}$ is the thermal 
size of the cloud, and hence the Kosterlitz-Thouless temperature is
$\sim \hbar^2 n_c/m \sim N^{1/2} \hbar \omega\sim T_c$.

An important feature of trapped gases is that the confining potential 
introduces a finite size of the sample, which sets a lower bound for the 
momentum of elementary excitations and reduces the phase fluctuations.
For this reason, at finite temperatures well below $T_c$ the phase fluctuations 
are small, and the equilibrium state is a true condensate. At intermediate 
temperatures $T<T_c$ the phase fluctuates on a distance scale smaller than 
the size of the gas sample, and one has a quasicondensate. 
Qualitatively, the character of the BEC state can be identified by 
comparing the size of the sample with the uniform-gas phase coherence 
length (\ref{lphi2D}) at the maximum density of the trapped gas.  
In this Lecture we present a more detailed analysis of the phase fluctuations 
and BEC character in weakly interacting trapped (quasi)2D gases. For simplicity
we assume that the coupling constant $g$ is independent of the gas density.
In the quasi2D case this requires the condition $l_0\gg |a|$ which leads to
$g$ given by Eq.(\ref{geff}). This condition is satisfied 
in recent experiments \cite{Ketlowd,Ing,Rudi,Eric2D}, 
where the quasi2D regime was reached for atomic Bose condensates.   

For finding the phase (and density) fluctuations in a trapped gas, one
should know the density profile and the spectrum of elementary excitations.
If the number of particles is very large and the chemical potential $\mu$
greatly exceeds the level spacing in the trap, the kinetic energy term in
Eq.~(\ref{overview:GPE}) is much smaller than the nonlinear
term and can be neglected. This approach is called the
Thomas-Fermi (TF) approximation, and in a harmonic trap the 
density profile takes the well-known parabolic shape:
\begin{equation}\label{overview:TF}
n=\frac{1}{g}\left(\mu-\frac{m}{2}\sum_i \omega_i^2r_i^2\right) \quad ,
\end{equation}
with $\omega_i$ and $r_i$ being the trap frequency and coordinate in the
$i$-th direction.
The dependence of the excitation spectrum on the trapping geometry has been
extensively studied for 3D TF condensates (see \cite{dalf99rmp} for review).
The spectrum and wavefunctions of low-energy excitations ($\varepsilon_\nu\ll\mu$) 
can be found analytically \cite{excStringari,excGora,excGraham}.  
For the 1D and 2D weakly interacting trapped Bose gases this has been done in 
\cite{stringari,Ho}.

We analyze the character of BEC in a harmonically trapped symmetric 
2D gas with repulsive interparticle interaction, relying on the
calculation of the single-particle correlation function \cite{Petrov2D}.
As well as in the uniform case, fluctuations of the density are provided
by excitations with wavelengths of the order of $1/l_c$ and in a similar
way one proves that they are small at temperatures $T\ll T_d$. Therefore,
the single-particle correlation function can be found by using the
field operator from Eq.(\ref{overview:fieldexpansion}), 
with coordinate-dependent mean
density $n({\bf r})$ and phase operator $\hat\phi_p({\bf r})$.  
Then the correlation function is given by Eq.(\ref{g1gen0}) generalized to
the trapped case: 
\begin{equation}     \label{matrix}
\langle\hat\Psi^{\dagger}({\bf r})\hat\Psi(0)\rangle=
\sqrt{n_0({\bf r})n_0(0)}\exp\left(- \frac{1}{2}\langle(\hat\phi_p(0)
-\hat\phi_p({\bf r}))^2\rangle \right) \quad ,
\end{equation}
where ${\bf r}=0$ at the trap center.

In the Thomas-Fermi regime, the radius of the gas sample is 
$R_{TF}=(2\mu/m\omega^2)^{1/2}$, and integration of Eq.(\ref{overview:TF}) 
over the spatial volume of the gas gives a relation between the chemical
potential and the number of particles:
\begin{equation}          \label{muN}
\mu=n_{0m}g=\sqrt{Nmg \over \pi\hbar^2} \hbar\omega \quad , 
\end{equation}
where $n_{m}=n(0)$ is the maximum density.  
It is worth noting that the ratio of the chemical potential to the
BEC transition temperature $T_c$ of an ideal gas is $\mu/T_c\approx
(mg/\pi\hbar^2)^{1/2}\ll 1$. 

For calculating the mean square fluctuations of the phase, we use
the (discrete) spectrum and wavefunctions of excitations with
energies $\varepsilon_{\nu}\ll\mu$, obtained relying on the method 
developed for Thomas-Fermi trapped condensates 
\cite{excStringari,excGora,excGraham,stringari,Ho}. 
At temperatures larger than
the chemical potential, for distances $r$ greatly exceeding the healing 
length $l_c$ at the trap center, we obtain a result similar to that in the uniform case:  
\begin{equation}     \label{ff}
\langle(\hat\phi_p(0)-\hat\phi_p({\bf r}))^2\rangle\equiv
(\delta\phi(r))^2\approx\left(\frac{mg}{\pi\hbar^2}\right)\frac{T}{\mu}
\ln\!\left(r\over l_c\right) \quad .
\end{equation} 

The character of the Bose-condensed state is determined by the 
phase fluctuations at distances $r\sim R_{TF}$. If they are small,
one has a true condensate, and for a large value of these fluctuations
the state is a quasicondensate. With logarithmic accuracy, 
from Eq.(\ref{ff}) we find
\begin{equation}   \label{fftr}
(\delta \phi(r\sim R_{TF}))^2\approx\left(\frac{mg}{2\pi\hbar^2}\right)
\frac{T}{\mu}\ln{N} \quad .
\end{equation}

In the case of a quasicondensate the phase coherence length $l_{\phi}$ is given
by an expression similar to that for uniform quasicondensates. For
$T>\mu$, equation (\ref{ff}) yields $l_{\phi}\approx l_c\exp(\pi\hbar^2n_m/mT)$.
As $l_{\phi}$ greatly exceeds the healing length, the quasicondensate has
the same Thomas-Fermi density profile as the true condensate. 
Correlation properties at distances smaller than $l_{\phi}$ 
and, in particular, local density correlators are also the same. 
Hence, one expects the same reduction of 
inelastic decay rates as in 3D condensates \cite{Kagan87}. 
However, the phase coherence properties of a quasicondensate are 
drastically different. We will have a detailed discussion of this
subject in the next Lecture for the case of 1D quasicondensates. 
 
From Eq.(\ref{fftr}) we see that in large gas samples one has 
more possibilities for obtaining the quasicondensate regime. 
In quasi2D trapped alkali gases one can expect the number of 
trapped atoms $N$ ranging from $10^4$ to $10^6$ and a value  
$\sim 10^{-2}$ or larger for the small parameter $mg/2\pi\hbar^2$. 
In the MIT sodium experiment \cite{Ketlowd} on achieving the quasi2D regime for
a trapped condensate, this small parameter was about $10^{-2}$ and 
the number of atoms was $\sim 10^5$. Therefore, even at temperatures
somewhat higher than $\mu$ the gas was in the regime of true BEC.
Note, however, that the coupling constant $g$ in the MIT experiment
\cite{Ketlowd} was rather small as the scattering length for sodium is only 
28 \AA. 
One can think of achieving the quasicondensate regime for the same
trapping parameters and $N$ as at MIT, by using Feshbach resonances and 
tuning the coupling constant to much larger values.

Presently, the largest obtained 2D Bose-condensed gas is the one of spin-polarized
atomic hydrogen on liquid helium surface \cite{Simo}. In this system the temperature
is about $100$ mK, the density exceeds $10^{13}$ cm$^{-3}$, and the number
of particles is $\sim 10^8$. Estimates show that the state should be 
a quasicondensate rather than a true condensate. However, this should still be 
proven experimentally.
 
{\it Problem}: Prove that to zero order in perturbation theory 
the coupling constant for the 2D Bose-condensed gas is $g=\hbar^2f/m$,
where $f$ is the scattering amplitude (as defined in Eq.(\ref{Asym})) 
at energy of the relative motion $E=2\mu$. {\it (Yu.E. Lozovik, 1971)}.

\section{{\it Lecture 3}. True and quasicondensates in 1D trapped gases} 

One-dimensional Bose systems at low temperature show a remarkable physics 
not encountered in 2D and 3D.
In particular, the 1D Bose gas with repulsive interparticle interaction
becomes more non-ideal with decreasing 1D density $n$ 
\cite{Girardeau,LiLi}.
The regime of a weakly interacting gas requires the correlation length
$l_c=\hbar/\sqrt{mng}$ to be much larger than the mean interparticle 
separation $1/n$.
For small $n$ or large interaction, where this condition is violated, the
gas acquires fermionic properties as the wavefunction strongly decreases at short 
interparticle distances \cite{Girardeau,LiLi}. 
In this case it is called a gas of impenetrable bosons or Tonks-Girardeau gas 
(cf. \cite{Tonks}).

Spatially uniform 1D Bose gases 
with repulsive interparticle interaction have been extensively studied in the
last decades. For the delta-functional interaction, 
Lieb and Liniger \cite{LiLi} have calculated the ground state energy
and the excitation spectrum which at low momenta turns out 
to be phonon-like. Generalizing the Lieb-Liniger approach, Yang and Yang
\cite{YY} have proved the analyticity of thermodynamical functions at any
finite  temperature $T$, which indicates the absence of a phase transition.  

However, at sufficiently low $T$ the correlation properties of
a 1D Bose gas are qualitatively different from classical high-$T$ properties.
In the regime of a weakly interacting gas the density 
fluctuations are suppressed \cite{KK}, whereas at finite $T$ the long-wave
fluctuations of the phase lead to an exponential decay of the single-particle
correlation function at large distances \cite{KK,RC}. A similar picture, with a
power-law decay of the density matrix, was found at $T=0$
\cite{Schwartz,Haldane}. Therefore, the Bose-Einstein condensate is absent at
any $T$, including $T=0$. These findings are consistent with
the Bogoliubov $k^{-2}$ theorem at finite $T$, and with a related treatment
at $T=0$ \cite{Pit}. 

Earlier studies of 1D Bose systems (see \cite{Popov} for review) allow us
to conclude that in uniform 1D  gases the decrease of temperature leads to a
continuous transformation of correlation properties from ideal-gas classical
to interaction/statistics dominated. In the weakly interacting regime at low 
$T$, where the density fluctuations are suppressed and the phase fluctuates 
on a distance scale much larger than the correlation length $l_c$, one can 
speak of a phase-fluctuating Bose-condensed state (quasicondensate).  
A 1D classical field model for calculating correlation functions in the 
conditions, where both the density and phase fluctuations are important, was 
developed in \cite{Sc} and for Bose gases in \cite{Castin}. The use of the 
Bogoliubov approach for describing 1D uniform (quasi)condensates is recently 
discussed in \cite{CastinMora} and is presented in the lectures of Yvan Castin.

In order to avoid misunderstanding, we should make a remark on the presence
of a quasicondensate in the uniform weakly interacting 1D Bose gas. It is 
sometimes stated that in this case the quasicondensate exists only at $T=0$,
and its presence is related to the power law decay of the single-particle
correlation function at large distances (see, for example, \cite{Griffin}).
However, at finite temperatures where this correlation function decays
exponentially, the phase coherence length $l_{\phi}$ can greatly exceed the 
healing length $l_c$. This is the case for small density fluctuations, and 
then the physical picture is similar to that in uniform 
finite-temperature 2D gases or zero-temperature 1D gases. The system can be 
divided into blocks of size 
$\tilde L$ satisfying the inequalities $l_c \ll \tilde L\ll l_{\phi}$, and inside 
each block one has a true condensate with a phase that is not correlated with 
phases in other blocks. In this sense, the term quasicondensate can also
be used in weakly interacting uniform 1D Bose gases at finite temperatures.
   
The 1D regime for trapped atomic condensates has been achieved in several 
experiments by tightly confining the motion of particles in two directions
\cite{Ketlowd,Schreck2001,Greiner2001}. In this Lecture we will discuss the various quantum degenerate states
that are present in finite-temperature 1D trapped Bose gases and focus 
attention on the case of a weakly interacting gas.   

\subsection{Weakly interacting regime in 1D}

We will discuss 1D Bose gases with repulsive short-range interaction between 
particles (coupling constant $g>0$). Present realizations of the 1D regime
imply particles in a cylindrical trap, which are tightly 
confined in the radial direction, with the confinement frequency 
$\omega_0$ greatly exceeding the mean-field interaction. Then, at 
sufficiently low $T$ the radial motion of particles is essentially 
``frozen'' and is governed by the ground-state wavefunction of the radial 
harmonic oscillator. If the radial extension of the wavefunction, 
$l_0=(\hbar/m\omega_0)^{1/2}$, is much larger than the radius of 
the interatomic potential $R_e$, the 
interaction between particles acquires a 3D character and is 
characterized by the 3D scattering length $a$. For this case, the coupling
constant has been found by Olshanii \cite{Olshanii}, and for $l_0\gg a$
it can also be obtained by averaging the 3D interaction over the radial
(harmonic oscillator) density profile:
\begin{equation}      \label{cc}
g=\frac{2\hbar^2}{m}\frac{a}{l_0^2}\quad ;\qquad l_0\gg a \quad .
\end{equation}
Statistical properties of such quasi1D samples are the same as those of a purely
1D system with the coupling constant given by Eq.(\ref{cc}).

As well as in the 2D case, we obtain the criterion of the weakly interacting 
regime at $T=0$ by comparing the interaction energy per particle, $I=ng$, with
a characteristic kinetic energy of particles at a mean separation ${\bar r}$ 
between them.
In the 1D case we have ${\bar r}\sim 1/n$, and this kinetic energy is $K\approx
\hbar^2n^2/m$.
In the weakly interacting regime, where the wavefunction of particles is not
influenced by the interaction at interparticle distances of the order of ${\bar r}$, 
one should have $I\ll K$. This leads to the criterion of the weakly interacting
regime 
\begin{equation} \label{small} 
\gamma=\frac{mg}{\hbar^2 n} \ll 1 \quad . 
\end{equation} 
From Eq.(\ref{small}) one really sees that in contrast to 2D and 3D gases, the 
1D Bose gas becomes more interacting with decreasing density. In the purely 
1D case this statement is valid as long as the mean interparticle separation $1/n$ 
remains much larger than the radius of interparticle interaction $R_e$. 
For quasi1D weakly interacting gases realized in present experiments, the density 
should still be such that the healing length $l_c=\hbar/\sqrt{mng}\gg l_0$, 
otherwise the gas leaves the 1D regime. Assuming $a \ll l_0$, this gives the
condition $na \ll 1$. 

The parameter $\gamma$ can also be interpreted as a ratio of the mean interparticle
separation to the characteristic interaction length for two particles, 
$r_g=\hbar^2/mg$. This quantity  
determines the distance scale on which the repulsion between the particles reduces 
their relative wavefunction. Under the condition $r_g\gg 1/n$, which is equivalent 
to Eq.(\ref{small}), this reduction is practically absent and, hence, the gas 
is weakly interacting.
In contrast, for $r_g\ll 1/n$ or, equivalently, $\gamma\gg 1$, the relative 
wavefunction is strongly reduced at distances smaller than $1/n$. The gas then 
enters the strongly interacting Tonks-Girardeau regime and acquires fermionic 
properties. 

For particles trapped in a harmonic (axial) potential $V(z)=m\omega^2z^2/2$,
one can introduce a complementary dimensionless quantity
\begin{equation}   \label{alpha}
\alpha=\frac{mg\,l}{\hbar^2} \quad ,
\end{equation}
where $l=\sqrt{\hbar/m\omega}$ is the amplitude of axial zero point oscillations.
The parameter $\alpha$ provides a relation between the interaction strength 
$g$ and the trap frequency $\omega$. Actually, it represents the ratio of $l$ to
the interaction length $r_g$. For $\alpha\ll 1$, the strongly interacting regime 
is not present at all as the relative motion of two particles on approach to each 
other is governed by their harmonic confinement rather than the interparticle 
interaction.  

\subsection{True BEC and diagram of states at $T=0$}

The nature of quantum degenerate states in the trapped 1D gas is strongly
influenced by the interparticle interaction and by the presence of the
trapping potential. The latter introduces a finite size of the sample
and provides a low-momentum cut-off for the phase and density fluctuations. 
This reduces the fluctuations compared to the uniform case.

We first discuss the case of $T=0$. For simplicity, we will consider a uniform
gas with a finite size $L$. We assume {\it a priori} that there is a true condensate 
with a density $n_0$ greatly exceeding the density of non-condensed particles $n'$.
Then, for a large value of $L$ we have the Bogoliubov excitation spectrum
given by Eq.(\ref{Bogsp}) and the excitation wavefunctions following from 
Eq.(\ref{fpmun}).
The density of non-condensed particles is given by an equation similar to 
Eq.(\ref{nprim}).
The difference is that now the lower limit of integration over the excitation momenta 
is non-zero. It is related to a finite size of the system and is equal to $\pi/L$.
We thus have
\begin{equation}     \label{T0n}
\frac{n'}{n_0}=\int_{\pi/L}^{\infty}\!\!\frac{dk}{\pi}\,\frac{v_k^2}{n_0}\quad ,
\end{equation}   
and a straightforward calculation yields
\begin{equation}    \label{T0nfin}
\frac{n'}{n_0}=\frac{\sqrt{\gamma}}{\pi}\ln\!\left(\frac{2L}{e\pi l_c}\right)\quad .
\end{equation}
 
For a realistic size $L$ of a trapped 1D Bose gas, the logarithmic factor in
Eq.(\ref{T0nfin}) does not exceed 10. Hence, for $\gamma\ll 1$ we have 
$n'\ll n_0$. This proves that fluctuations are small and justifies our
initial assumption on the presence of a true BEC in a weakly interacting
trapped Bose gas at $T=0$. An analysis of vacuum fluctuations for 
1D Bose gases in harmonic traps is given in \cite{Ho}.

Thus, in a harmonically trapped 1D Bose gas at $T=0$ we have a true
condensate with the density profile determined by the Gross-Pitaevskii equation
(\ref{GP}). In the Thomas-Fermi (TF) regime, where the chemical potential is
$\mu\gg\hbar\omega$, the kinetic energy term can be omitted and we have a  
parabolic density distribution typical for condensates in harmonic traps:
\begin{equation}   \label{1Ddistr}
n_0(z)=n_{\rm 0m}\left(1-\frac{z^2}{R_{TF}^2}\right)
\end{equation}
for $-R_{TF}\leq z\leq R_{TF}$, and is equal to zero otherwise.
The (half)size of the condensate is $R_{TF}=(2\mu/m\omega^2)^{1/2}$, and
the maximum  density is $n_0(0)=n_{\rm 0m}=\mu/g$. Integrating 
Eq.(\ref{1Ddistr}) over $dz$ we obtain a relation between 
the chemical potential and the number of particles $N$:
\begin{equation}      \label{muN1D}   
\mu=\hbar\omega\left(3N\alpha \over 4\sqrt2\right)^{2/3} \quad ,  
\end{equation}
where the parameter $\alpha$ is given by Eq.(\ref{alpha}). 

For $\alpha\gg 1$ we are always in the TF regime. In this case, 
Eq.(\ref{small}) requires a sufficiently large number of particles: 
\begin{equation}    \label{TF}
N\gg N_*=\alpha^2 \quad ,
\end{equation}
which reflects the fact that the weakly interacting regime requires sufficiently
large densities. For $\alpha\ll 1$ the criterion (\ref{small}) of a weakly 
interacting gas is satisfied at any $N$. The condensate is in the TF regime 
if $N\gg\alpha^{-1}$.
In the opposite limit the mean-field interaction is much smaller than the level
spacing in the trap $\hbar\omega$. Hence, one has a macroscopic occupation of the
ground state of the trap, i.e. there is an ideal gas condensate with a Gaussian 
density profile.

If $\alpha\gg 1$ and $N\ll N_*$, then the trapped 1D gas is in the strongly
interacting Tonks-Girardeau regime.
The one-to-one mapping of this system to a gas of free
fermions \cite{Girardeau} ensures the fermionic spectrum and density
profile. The chemical potential is equal to $N\hbar\omega$, and the density 
distribution is $n(z)=(\sqrt{2N}/\pi l)\sqrt{1-(z/R)^2}$, where the size of the
cloud is $R=\sqrt{2N}l$. Note that this profile is different from
both the profile of the zero-temperature condensate and the classical
distribution of particles. 
Correlation properties of strongly interacting 1D Bose gases will be discussed in 
Lecture 4.

In Fig.5, we present the diagram of states for the zero-temperature trapped
1D Bose gas in terms of $N$ and $\alpha$. This diagram clearly shows the presence
of three states as discussed above: Thomas-Fermi BEC, condensate with a Gaussian 
density profile, and the Tonks-Girardeau gas. 

\begin{figure}
\centerline{\includegraphics[width=.7\textwidth]{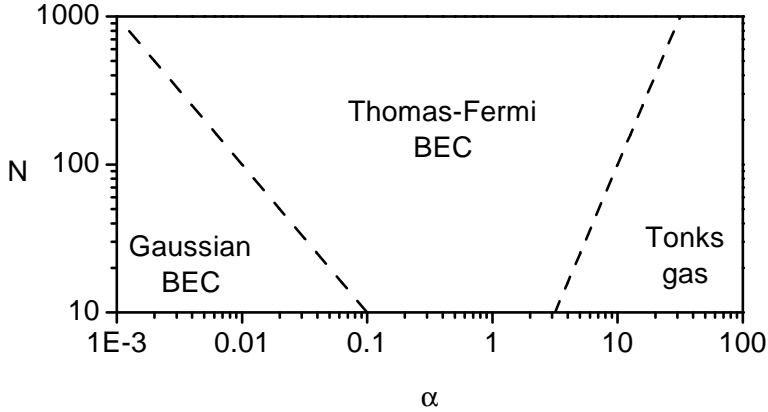}}
  \caption{Diagram of states for a trapped 1D gas at $T=0$.}
\label{Fig31}
\end{figure}

\subsection{Regimes of quantum degeneracy at finite temperature}

The characteristic temperature of quantum degeneracy for a trapped 1D Bose
gas is $T_d\approx N\hbar\omega$. In the weakly interacting regime, the character
of finite-temperature Bose-condensed states at $T\ll T_d$ 
is determined by thermal fluctuations.
Before describing these states, we briefly analyze a cross-over to the BEC regime 
in 1D trapped gases. 

As we discussed in Lecture 1, for an ideal gas one has 
a sharp cross-over to BEC. Ketterle and van Druten \cite{vDK}
found that the decrease of $T$ to below $T_c=N\hbar\omega/\ln(2N)$
strongly increases the  population of the ground state, which rapidly becomes
macroscopic. This sharp cross-over originates from the discrete structure of
the trap levels. However, the presence of the interparticle  interaction
changes the picture drastically. One can distinguish between the (lowest)
trap levels only if the interaction between particles occupying a particular
level is much smaller than the level spacing. Otherwise the interparticle
interaction smears out the discrete structure of the levels. For $T$ close to
$T_c$ the occupation of the ground state is $\sim T_c/\hbar\omega\approx
N/\ln(2N)$ \cite{vDK} and, hence, the mean-field interaction  between the
particles in this state (per particle) is $Ng/l\ln(2N)$. The sharp BEC
cross-over requires this quantity to be much smaller than $\hbar\omega$, and
we arrive at the condition $N/\ln(2N)\ll \alpha^{-1}$. For a realistic number
of particles ($N\sim 10^3 - 10^4$) this is practically equivalent to
the condition $N\ll \alpha^{-1}$ at which one has the ideal gas Gaussian 
condensate. 

As we see, the sharp BEC cross-over requires small $\alpha$. 
For possible realizations of 1D gases, using
the coupling constant (\ref{cc}) for $l_0\gg a$, we obtain
$\alpha=2al/l_0^2$. Then, even for the ratio $l/l_0\sim 10$ and
moderate radial confinement with $l_0\sim 1~\mu$m, we have $\alpha\sim
0.1$ for Rb atoms ($a\approx 50$ \AA). Clearly, for a reasonably large number
of particles the (sharp) cross-over condition $N \ll \alpha^{-1}$ can only be fulfilled
at extremely small interparticle interaction. One can think of reducing $a$  to
below $1$\AA$\,$ and achieving $\alpha<10^{-3}$ by using Feshbach resonances.
In this case one can expect the sharp BEC cross-over already for $N\sim 10^3$.

Otherwise, similarly to the uniform 1D case, the decrease of temperature to below
$T_d$ continuously transforms a classical 1D gas to the regime of quantum degeneracy.
If the condition (\ref{small}) is satisfied, one has a weakly interacting gas which
at $T=0$ becomes a true TF condensate. To understand the nature of a 
Bose-condensed state at finite $T \ll T_d$, we analyze the behavior of the
single-particle correlation function by calculating the
fluctuations of the density and phase \cite{Petrov1D}.  

We assume {\it a priori} that the weakly interacting trapped 1D Bose gas is 
characterized by small density fluctuations and justify this assumption later.
Then the operators of the density and phase fluctuations are given by 
Eqs.~(\ref{nop}) and (\ref{phiop}), and the spectrum and wavefunctions of elementary 
excitations follow from Eqs.~(\ref{BdG+}) and (\ref{BdG-}) in which the 
density distribution is the one of the zero-temperature condensate.
It has already been shown in section 3.2 that vacuum fluctuations of the density and
phase are small. Therefore, we now discuss only thermal fluctuations.
Using Eq.(\ref{muN1D}) we find that $\mu/T_d\sim (\alpha^2/N)^{1/3}\ll 1$.
Accordingly, our analysis will include both cases, $T\ll\mu$ and $T\gg\mu$.

Fluctuations coming from the high-energy part of the excitations 
($\varepsilon_{\nu}>\mu$) are small. Similarly to the uniform case discussed
in Lecture 2, this part can be viewed as an ideal thermal gas of particles, with 
chemical potential equal to $-\mu$. One then finds that the number of particles in
the high-energy part becomes exponentially small for $T\ll\mu$, and for $T\gg\mu$
it is $\sim (T/T_d)N\ll N$. 
We therefore confine ourselves to fluctuations coming from low-energy
excitations ($\varepsilon_{\nu}<\mu$).

The solution of Eqs.~(\ref{BdG+}) and (\ref{BdG-}) for these excitations gives the
spectrum $\varepsilon_j=\hbar\omega\sqrt{j(j+1)}$ \cite{stringari,Ho} and wavefunctions
\begin{equation}   \label{fpm1Dtrap}
f_j^{\pm}=\left(\frac{j+1/2}{R_{TF}}\right)^{1/2}
\left[\frac{2\mu}{\varepsilon_j}(1-x^2)\right]^{1/2}P_j(x) \quad ,
\end{equation}
where $j$ is a positive integer, $P_j$ are Legendre polynomials, and
$x=z/R_{TF}$. Using Eqs.~(\ref{nop}) and (\ref{fpm1Dtrap}), for the
mean square (thermal) fluctuations of the density we have
\begin{equation}    \label{dfl1Dtrap}  
\langle(\delta\hat n(z)-\delta\hat n(z'))^2\rangle_T
\equiv\langle\delta n^2_{zz'}\rangle_T=
\sum_{j=1}^{\varepsilon_j<\mu}\frac{\varepsilon_j n_{0m}(j+1/2)}
{\mu R_{TF}}(P_j(x)-P_j(x'))^2 N_j \quad ,
\end{equation}
with $N_j=[\exp(\varepsilon_j/T)-1]$ being the occupation numbers for the
excitations. Assuming $T\gg\hbar\omega$, the main contribution to the 
density fluctuations comes from quasiclassical excitations ($j\gg 1$). 
These fluctuations are largest on a distance scale $|z-z'|$ greatly 
exceeding the de Broglie wavelength of the excitations. In this case   
we obtain
\begin{equation}    \label{dfl1Dtrapfin}  
\frac{\langle\delta n^2_{zz'}\rangle_T}{n_{0m}^2}=\frac{T}{T_d}
{\rm min}\{(T/\mu),1\} \quad .
\end{equation}
Thus, we see that the density fluctuations are small at any temperature 
$T\ll T_d$.

The result of Eq.(\ref{dfl1Dtrapfin}) looks quite different from what one
finds for a uniform 1D Bose gas. In the latter case, using the Bogoliubov 
spectrum (\ref{Bogsp}) and excitations wavefunctions (\ref{fpmun}), for $T\gg\mu$ 
we have $\langle\delta n^2_{zz'}\rangle_T\sim (T/\sqrt{\mu T_d})n^2$.
The difference from Eq.(\ref{dfl1Dtrapfin}) is related to the fact that 
in the trapped case the density profile
shrinks when decreasing temperature well below the global temperature of
quantum degeneracy $T_d=N\hbar\omega$. Therefore, at $T\ll T_d$ the local
value of the degeneracy temperature in the center of the trap is 
$T_{0}\sim 2\pi\hbar^2n_{0m}^2/m\sim T_d^2/\mu\gg T_d$. Expressing $T_d$
in Eq.(\ref{dfl1Dtrapfin}) through $T_{0}$ we restore qualitatively the
uniform-gas result.    

As the low-energy fluctuations of the density are suppressed and high-energy
fluctuations are also small, for finding the single-particle correlation 
function we may use the field operator in the form 
(\ref{overview:fieldexpansion}): $\hat\Psi(z)=
\sqrt{n_0(z)}\exp(i\hat\phi_p(z))$, where the operator $\hat\phi_z$ represents the 
low-energy part of the phase fluctuations. Then, similarly to the 2D case, 
the correlation function is expressed through the mean square fluctuations 
of the phase:
\begin{equation}    \label{g11Dtrap}
g_1(z,z')\equiv\langle\hat\Psi^{\dagger}(z')\hat\Psi(z)\rangle= 
\sqrt{n_0(z)n_0(z')}\exp\left(-\frac{1}{2}\langle\delta\hat\phi^2_{zz'}\rangle\right) \quad ,
\end{equation}
where $\delta\hat\phi_{zz'}=\hat\phi_p(z)-\hat\phi_p(z')$. On the basis of
Eqs.~(\ref{phiop}) and (\ref{BdG+}), for the thermal phase fluctuations
we have
\begin{equation}    \label{phifl1Dtrap}
\langle \delta\hat\phi^2_{zz'}\rangle_T=
\sum_{j=1}^{\varepsilon_j<\mu}\frac{\mu (j+1/2)}
{\varepsilon_j n_{0m}R_{TF}}(P_j(x)-P_j(x'))^2 N_j \quad .
\end{equation}
In contrast to the 2D gas, thermal fluctuations of
the phase in 1D are mostly provided by the contribution of the
lowest excitations. A direct calculation of Eq.(\ref{phifl1Dtrap}) yields 
\begin{equation} \label{phifin}
\langle\delta\hat\phi_{zz'}^2\rangle_T=\frac{4T\mu}{3T_d\hbar\omega}\left|\ln
\left[\frac{(1-x')}{(1+x')}\frac{(1+x)}{(1-x)}\right]\right| \quad .
\end{equation}
For $z$ and $z'$ close to the trap center, the logarithmic term in
Eq.(\ref{phifin}) is equal to $2|z-z'|/R_{TF}\ll 1$. Otherwise, it is 
of order unity, except for the region close to the Thomas-Fermi border
of the density distribution.

The temperature $T_{\phi}$ at which the quantity 
$\langle\delta\hat\phi_{zz'}^2\rangle$ is of order unity on a
distance scale $|z-z'|\sim R_{TF}$, is given by 
\begin{equation} \label{Tph}
T_{\phi}=T_d \frac{\hbar\omega}{\mu} \quad .
\end{equation}   
Thus, for $T\ll T_{\phi}$ both the density and phase fluctuations are suppressed, and 
there is a true condensate. The condition (\ref{TF}) always provides the ratio
$T_{\phi}/\hbar\omega\approx (4N/\alpha^2)^{1/3}\gg 1$. 

In the temperature range, where $T_d\gg T\gg T_{\phi}$, the density
fluctuations are suppressed, but  the phase fluctuates on a distance scale
$l_{\phi}\approx R_{TF}(T_{\phi}/T)\ll R_{TF}$. Hence, 
similarly to the 2D case, we have a condensate with fluctuating phase  
(quasicondensate). The phase coherence
length $l_{\phi}$ greatly exceeds the correlation length: $l_{\phi}\approx
l_c(T_d/T)\gg l_c$. Therefore, the quasicondensate has the same
density profile and local correlation properties as the true condensate.

\begin{figure}
\centerline{\includegraphics[width=.7\textwidth]{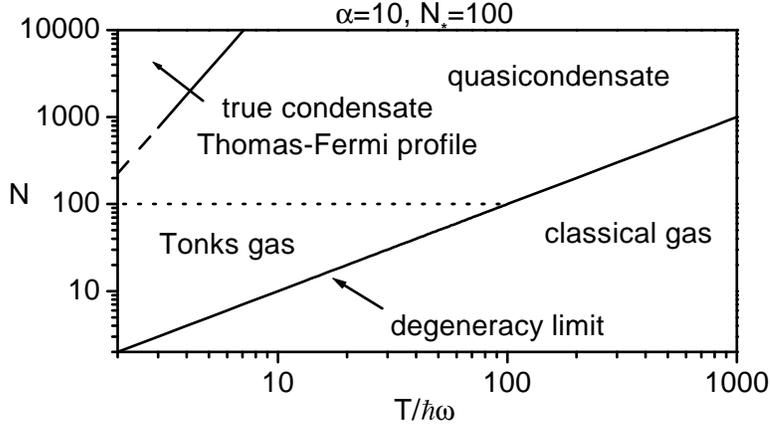}}
  \caption{Diagram of states for a finite-temperature trapped 1D Bose gas.}
\label{Fig32}
\end{figure}

In Fig.~\ref{Fig32}, we present the state diagram for the finite-temperature 
trapped 1D gas at $\alpha=10$ ($N_*=100$). One clearly sees three 
quantum degenerate regimes:
the BEC regimes of a quasicondensate and true condensate, and the 
regime of a trapped Tonks gas \cite{Petrov1D}.
For $N\gg N_*$, the decrease of temperature to below $T_d$ leads to 
the appearance of a quasicondensate which at $T<T_{\phi}$ turns to the 
true condensate. In the $T-N$ plane the approximate border line between 
the two BEC regimes is determined by the equation
$(T/\hbar\omega)=(32N/9N_*)^{1/3}$. For $N<N_*$ the system can be 
regarded as a trapped Tonks gas. 

\begin{figure}
\centerline{\includegraphics[width=.7\textwidth]{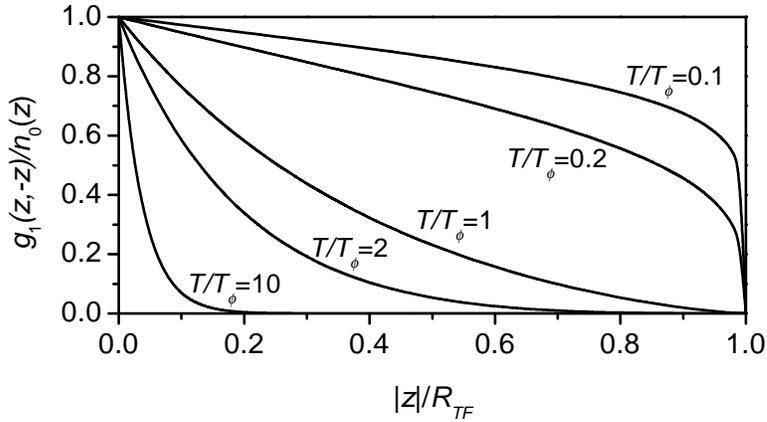}}
\caption{Normalized single-particle correlation function 
$g_1(z,-z)/n_0(z)$ versus $|z|/R_{TF}$.}
\label{Fig33}
\end{figure}

A cross-over from one regime to another is always smooth. The absence of
a sharp transition from true to quasiBEC is seen from the behavior of the
single-particle correlation function. Using Eqs.~(\ref{g11Dtrap}) and 
(\ref{phifin}) and omitting small vacuum fluctuations, for the normalized 
correlation function at $z'=-z$ we obtain
\begin{equation}     \label{g1Tpower}
\frac{g_1(z,-z)}{n_0(z)}=\left(\frac{1-|z|/R_{TF}}
{1+|z|/R_{TF}}\right)^{4T/3T_{\phi}} \quad .
\end{equation}
This function is displayed in Fig.~\ref{Fig33} for various ratios $T/T_{\phi}$.
In particular, we see that the full phase coherence requires temperatures
well below $T_{\phi}$.

Present facilities allow one to achieve both the true and quasicondensate
regimes for the 1D trapped Bose gas. For example, in the case of $10^3$ 
rubidium atoms trapped in a cylindrical harmonic potential with axial
frequency $\omega\sim 1$ Hz, we have the temperature of quantum degeneracy
$T_d\sim 100$ nK. Then, if the tight radial confinement providing the 1D
regime is made at a frequency $\omega_0\sim 100$ Hz, the characteristic 
temperature $T_{\phi}$ is about $10$ nK. 

The phase fluctuations lead to a drastic difference of the phase coherence 
properties of a quasicondensate from those of a true condensate. 
This can be understood from a gedanken `juggling'  experiment
similar to those with 3D condensates at NIST and Munich \cite{Phi,Ess}. 
Small clouds of atoms are ejected from the main cloud by stimulated Raman
or RF transitions. Observing the interference between two clouds, simultaneously 
ejected from different parts of the sample, allows  the reconstruction of the 
spatial phase coherence properties. Repeatedly juggling clouds of a
small volume $\Omega$ from points $z$ and $z'$ of the 1D sample, for equal time
of flight to the detector we have the averaged detection signal 
$I=\Omega[n_0(z)+n_0(z')+2\langle\hat\psi^{\dagger}(z)\psi(z')\rangle]$ and
thus measure directly the single-particle correlation function 
$g_1(z,z')=\langle\hat\psi^{\dagger}(z)\psi(z')\rangle$.   

In the regime of a weakly interacting gas, at $T\ll T_{\phi}$ the phase
fluctuations are small and one has a true condensate. In this case, for
$z'=-z$ we have $\langle\hat\psi^{\dagger}(z)\hat\psi(z')\rangle=n_0(z)$ and
$I=4\Omega n_0(z)$, which shows a pronounced
interference effect. The detected signal is
twice as large as the number of atoms in the ejected clouds. 
The phase fluctuations grow with $T$ and for $T>T_{\phi}$, where the true
condensate turns into a quasicondensate, the detection signal decreases as
described by $g_1(z,-z)$ from Eq.(\ref{g1Tpower}). For $T\gg T_{\phi}$ the phase
fluctuations completely destroy the interference between the two ejected
clouds, and $I=2\Omega n_0(z)$. 

The time-dependent single-particle correlation function $\langle\hat\Psi^{\dagger}
(z,t)\hat\Psi(z',t')\rangle$ has been calculated in \cite{Griffin} and it has been
shown how this correlation function can be measured in two-photon Rahman outcoupling 
experiments.

\subsection{Phase coherence in 3D elongated condensates}

The one-dimensional character of thermal phase fluctuations is also present
in a very elongated 3D Bose gas, which leads to the appearance
of a quasicondensate in this system. This phenomenon, predicted 
\cite{Petrovelongated} and observed \cite{HannExp1,HannExp2,Orsay1,Orsay2,Arlt} 
in recent studies, is described by the following physical 
picture. Excitations of elongated condensates can be divided into two groups:
``low energy'' axial excitations with energies
$\varepsilon_{\nu}<\hbar\omega_0$, and ``high energy'' excitations with
$\varepsilon_{\nu}>\hbar\omega_0$. The latter have 3D
character as their wavelengths are smaller than the radial size $R_{\perp}$. 
Therefore, as in ordinary 3D condensates, these excitations
can only provide small phase fluctuations. The ``low-energy'' axial excitations
have wavelengths larger than $R_{\perp}$ and exhibit a pronounced 1D behavior. 
Just these excitations give the most important contribution to the long-wave axial
fluctuations of the phase. 

A detailed description of fluctuations in elongated 3D condensates is given in
\cite{Petrovelongated}, and in this Lecture we only briefly outline the results. 
We consider a cylindrical Thomas-Fermi condensate with an axial size $R_{TF}$ 
greatly exceeding
the radial size $R_{\perp}$, and assume that most particles are Bose-condensed.
The axial thermal fluctuations of the phase, with wavelengths larger than 
$R_{\perp}$, are similar to the fluctuations in the 1D case. In particular, 
one finds that the mean square fluctuations on a distance scale $\sim R_{TF}$ 
are approximately equal to $T/T_{\phi}$, where the characteristic temperature 
$T_{\phi}$ is given by  
\begin{equation} \label{Tphi}
T_{\phi}=15\frac{(\hbar\omega)^2}{32\mu}N \quad ,
\end{equation}   
with $N$ being the number of particles, and $\omega$ the axial trap
frequency. This is qualitatively the same as the result of Eq.(\ref{Tph})
where one substitutes $T_d=N\hbar\omega$. Accordingly, the phase coherence
length is again determined by the relation $l_{\phi}=R_{TF}(T_{\phi}/T)$.
It is much larger than the healing length: 
$l_{\phi}/l_c\approx (T_c/T)(T_c/\hbar\omega_0)^2\gg 1$, with $T_c\approx
(N\omega_0^2\omega)^{1/3}$ being the BEC transition temperature, and
$\omega_0$ the radial confinement frequency ($\omega_0\gg\omega$).

We thus see that the situation is quite similar to that in 1D trapped gases.
One has a continuous transformation of a quasicondensate into a true BEC 
when decreasing temperature to below $T_{\phi}$.  

\begin{figure}
\centerline{\includegraphics[width=1.\textwidth]{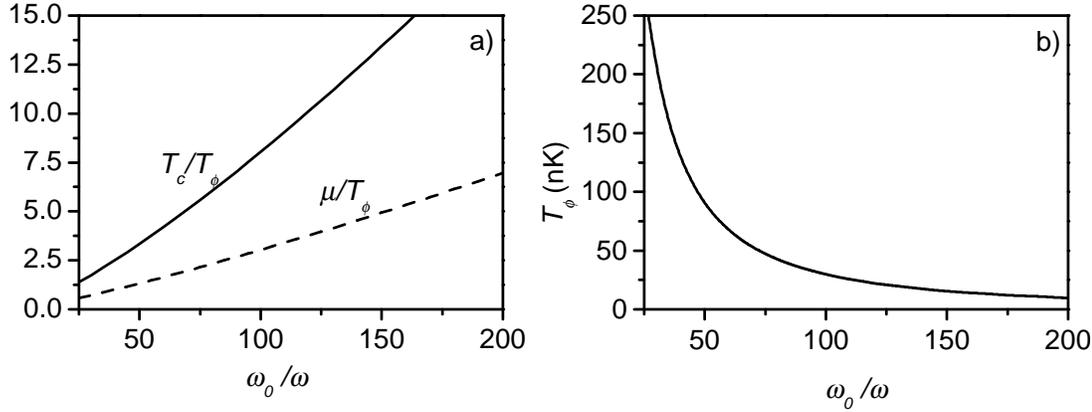}}
  \caption{The ratios $T_c/T_{\phi}$ and $\mu/T_{\phi}$
in (a) and the temperature $T_{\phi}$ in (b), versus the aspect ratio
$\omega_{0}/\omega$ for trapped Rb condensates with
$N=10^5$ and $\omega_{0} =500$ Hz.}
\label{Fig34}
\end{figure}

Most important is the dependence of the phase fluctuations on the aspect 
ratio of the cloud $\omega_0/\omega$. Fig.~\ref{Fig34} shows 
the ratios $T_c/T_{\phi}$, $\mu/T_{\phi}$, and the temperature
$T_{\phi}$ as functions of $\omega_0/\omega$ for Thomas-Fermi
rubidium condensates at $N=10^5$ and $\omega_0=500$ Hz.
From these results we see
that 3D quasicondensates can be obtained in elongated geometries with 
$\omega_0/\omega\gtrsim 50$.

The phase fluctuations are very sensitive to temperature. From
Fig.8 we see that one can have $T_{\phi}/T_c<0.1$,
and the phase fluctuations are still significant at $T<\mu$, where only a 
tiny indiscernible thermal cloud is present. 
This suggests a principle for thermometry of 3D Bose-condensed gases with 
indiscernible thermal clouds. If the sample is not an elongated quasicondensate
by itself, it is first transformed to this state by adiabatically increasing
the aspect ratio $\omega_{\rho}/\omega_z$. This does not change the ratio
$T/T_c$ as long as the condensate remains in the 3D Thomas-Fermi regime.
Second, the phase coherence length $l_{\phi}$ or the single-particle
correlation function are measured. These quantities depend on temperature if
the latter is of the order of $T_{\phi}$ or larger. One thus can measure the
ratio $T/T_c$ for the initial cloud,  which is as small as the ratio
$T_{\phi}/T_c$ for the elongated cloud. 

\begin{figure}
\centerline{\includegraphics[width=.7\textwidth]{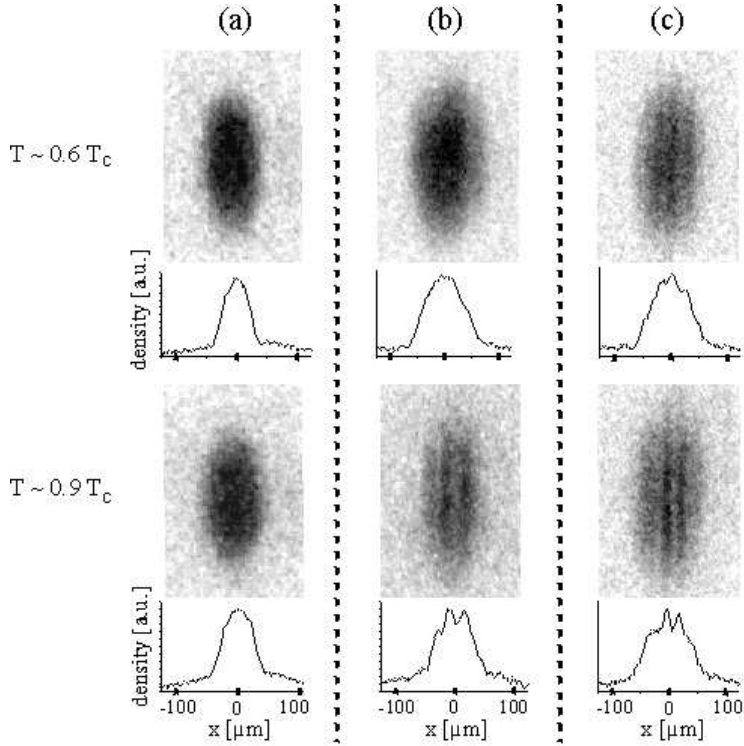}}
  \caption{Absorption images and corresponding density profiles of
BECs after $25\, $ms time-of-flight in the Hannover experiment
for aspect ratios [$\omega_{0}/\omega = 10$ (a), 26 (b), 51 (c)].}
\label{Fig35}
\end{figure}

Pronounced phase fluctuations have been first observed in Hannover experiments
with very elongated cylindrical 3D condensates of up to $10^5$ rubidium atoms
\cite{HannExp1,HannExp2}. The expanding cloud released from the trap was imaged after
25 ms of time of flight and the images showed clear modulations of the density
(stripes) in the axial direction (see Fig.~\ref{Fig35}). The physical reason for the
appearance of stripes is the following. In a trap the density distribution
does not feel the presence of the phase fluctuations, since the mean-field
interparticle interaction prevents the transformation of local velocity fields
provided by the phase fluctuations into modulations of the density. After
switching off the trap, the cloud rapidly expands in the radial direction,
whereas the axial phase fluctuations remain unaffected. As the mean-field
interaction drops to almost zero, the axial velocity fields are then converted 
into the density distribution.

The mean square modulations of the density in the expanding cloud provide a
measure of the phase fluctuations in the initial trapped condensate. A direct
relation between these quantities has been established from analytical and
numerical solutions of the Gross-Pitaevskii equation for the expanding cloud,
with explicitly included initial fluctuations of the phase \cite{HannExp1}.
The obtained phase coherence length was inversely proportional to $T$, in
agreement with theory, and for most measurements it was smaller than the axial
size of the trapped Thomas-Fermi cloud. This implies that the measurements
were performed in the regime of quasicondensation.   

The properties of quasicondensates and the phase coherence length were
measured directly in Bragg spectroscopy experiments with elongated rubidium
BECs at Orsay \cite{Orsay1}. In this type of experiment one measures the
momentum distribution of particles in the trapped gas. The Orsay studies 
\cite{Orsay1} find a Lorentzian momentum distribution
characteristic of quasicondensates with axially fluctuating phase
\cite{Orsay2}, whereas a true condensate has a Gaussian distribution. The
width of the Lorentzian momentum distribution is related to the phase
coherence length at the trap center. These investigations are described in
detail in the lecture of Philippe Bouyer. The phase coherence length has also 
been found in Hannover experiments \cite{Arlt} from the measurement of the 
intensity correlation function of two interfering spatially displaced copies
of phase-fluctuating condensates.   

It is important to emphasize that the measurement of phase correlations 
will allow one to study the evolution of phase coherence in the course of the
formation of a condensate out of a non-equilibrium thermal cloud. This problem 
has a rich physics. For example, recent experiments on the formation kinetics
of trapped condensates \cite{Jook} indicate the appearance of non-equilibrium
quasicondensates slowly evolving towards the equilibrium state.

{\it Problem}: Calculate the momentum distribution for the weakly interacting
1D Bose gas in a rectangular box of size $L$ at temperatures much smaller than 
the temperature of quantum degeneracy. Describe how the momentum distribution 
changes when the gas transforms from the true to quasicondensate.  

\section{{\it Lecture 4}. Correlations in strongly interacting 1D Bose gases}

In Lectures 2 and 3, we saw that the reduction of spatial dimensionality in trapped
Bose gases increases the number of possible quantum degenerate states. 
In 3D we have (true) Bose-Einstein condensates and they were extensively
studied in a large variety of experiments during last years. In the 2D case,
there are two types of Bose-condensed states: true condensates and
quasicondensates. In 1D, besides these BEC states, one can have a trapped
Tonks gas which should exhibit some of the fermionic properties. 
Static and dynamical properties of strongly interacting trapped Bose gases
are discussed in the lectures of Maxim Olshanii and Sandro Stringari.

In this Lecture we discuss correlation properties of strongly 
interacting Bose gases, which are drastically different from those in the 
weakly interacting regime. This is especially interesting in view of a 
rapid progress of experimental studies which have already reached an 
intermediate regime between the weakly and strongly interacting ones 
\cite{Ess03,Phi03}.

Beyond the weakly interacting regime one can no longer use the mean-field
Bogoliubov approach. Nevertheless, long-wave properties are generic and can
be studied relying on the quantum hydrodynamic approach developed by Haldane
\cite{Haldane} and widely used for uniform 1D systems (see \cite{Caz} for a recent 
overview). We will employ this approach 
for studying long-distance phase coherence in 1D trapped Bose gases \cite{Gangardt1}.

The problem of short-range correlations requires more sophisticated approaches.
The term "short-range" is used in the sense that
the distance is of the order of or smaller than the characteristic correlation
length of the gas, which for repulsive interaction is $l_c=\hbar/\sqrt{m\mu}$  
with $\mu$ being the chemical potential. However, the distance is still much
larger than the radius of interatomic potential, $R_e$. Then, in the purely
1D case such short-range correlations may be investigated by using the 
Lieb-Liniger model which assumes a delta-functional interaction between the atoms.

The Lieb-Liniger model for a uniform  system is exactly solvable by using the Bethe 
Ansatz \cite{Exact1D}. Thermodynamic functions for this model at zero and finite 
temperatures have been found by Lieb and Liniger \cite{LiLi} and 
by Yang and Yang \cite{YY}. On the other hand,  
the problem of correlation properties is far from being completely resolved, 
except for some correlation functions in the limiting cases of weak and 
strong interactions. For example, the case of infinitely strong
interactions is to a certain extent  equivalent to that of
free fermions and the interactions play the role of Pauli
principle \cite{Girardeau}. In this limit, any
correlation function of the density is given by the corresponding
expression for fermions \cite{Mehta}. 
The expressions for the one-body and two-body correlations for an arbitrary
interaction strength were obtained by using the Inverse
Scattering Method \cite{KorepinBook}. However, closed analytical results can
be found only as perturbative expansions in the limiting cases of strong and
weak interactions \cite{Lenard,Vaidya,Jimbo,Korepin,Creamer}.

We will demonstrate the use of the Lieb-Liniger model for finding local 2-body
and 3-body correlations, that is the correlations at distances much smaller
than the correlation length $l_c$ \cite{Gangardt1,Gangardt2,Gangardt3}. 
These correlations are responsible for the stability of the gas with regard to 
intrinsic inelastic processes, such as 3-body recombination.
 
\subsection{Lieb-Liniger model for trapped 1D gases}
 
In the case of 1D trapped gases, particles undergo zero point oscillations
in two (radial) tightly confined directions and there is a question
to which extent one can use the 1D Lieb-Liniger model for describing the
system. We will consider repulsive interaction between particles.
The 1D regime is realized if the amplitude of radial zero
point oscillations $l_0=\sqrt{\hbar/m\omega_0}$ is much smaller than the
(axial) correlation length $l_c=\hbar/\sqrt{m\mu}$, where
$\omega_0$ is the frequency of the radial confinement and 
$\mu$ is the chemical potential of the 1D system. This is equivalent to the 
condition $\mu\ll \hbar\omega_0$. One then
has a 1D system of bosons interacting with each other via a
short-range potential characterized by an effective coupling constant $g>0$.
This constant is expressed through the 3D scattering length $a$ 
\cite{Olshanii} and for $l_0\gg a$ is given by Eq.(\ref{cc}). Accordingly,
the 1D interaction (scattering) length is $r_g\sim l_0^2/a\gg a$. 
In the weakly interacting regime, the chemical potential is $\mu\approx gn$, 
and the condition $l_c\gg l_0$ leads to the inequality $na\ll 1$. In the 
Tonks-Girardeau regime the correlation length $l_c\sim 1/n$, and one 
should have $nl_0\ll 1$. We thus see that irrespective of the interaction 
strength, it is sufficient to satisfy the inequalities
\begin{equation}
\label{ineq}
a \ll l_0 \ll \frac{1}{n} \quad .
\end{equation}
Then the 1D regime is reached and correlation properties of the system can be
analyzed on the basis of the 1D Lieb-Liniger model, which in the absence of
an axial trapping potential is described by the Hamiltonian
\begin{equation}      \label{LLham}
H=\sum_{j=1}^N -\frac{\hbar^2}{2m}\partial^2_{x_j}+g\sum_{i<j} 
\delta(x_i-x_j) \quad ,
\end{equation}
with $N$ being the number of particles. From the Hamiltonian (\ref{LLham}) 
one easily finds that the ratio of the characteristic kinetic energy of 
particles at the mean interparticle separation to the interaction energy 
per particle, is given by the parameter
$\gamma=mg/\hbar^2n$ introduced in Lecture 3 in Eq.(\ref{small}). In the
strongly interacting regime we have $\gamma\gg 1$.       

At zero temperature the energy of the system $E_0$ can be written as
\begin{equation}     \label{E0}
\frac{E_0}{N}=\frac{\hbar^2n^2}{2m}e(\gamma) \quad ,
\end{equation}
where $\gamma$ is given by Eq.(\ref{small}) but is not necessarily small, 
and the function $e(\gamma)$ follows 
from the Bethe Ansatz solution found by Lieb and Liniger \cite{LiLi}. 
This function is calculated numerically for any value of $\gamma$ 
\cite{LiLi,Dunko,Ciara1}.
In the weakly interacting regime of a quasicondensate ($\gamma\ll 1$) we have
\begin{equation}    \label{eweak}
e(\gamma)=\gamma-\frac{4}{3\sqrt{\pi}}\gamma^{3/2} \quad ; \qquad \gamma\ll 1 \quad ,
\end{equation}
which coincides with the result of the Bogoliubov approach. For the strongly
interacting Tonks-Girardeau regime ($\gamma\gg 1$) the function $e(\gamma)$
is given by
\begin{equation}    \label{estrong}
e(\gamma)=\frac{\pi^2}{3}\left(1-\frac{4}{\gamma}\right) \quad ; \qquad \gamma \gg 1 \quad .
\end{equation}
The expression for the chemical potential follows immediately from 
Eq.(\ref{E0}):
\begin{equation}            \label{muLL}
\mu=\frac{\partial E_0}{\partial N}=\frac{\hbar^2 n^2}{2m}
\left(3e(\gamma)-\gamma\frac{d e(\gamma)}{d\gamma}\right) \quad .
\end{equation}   
Accordingly, in the weakly interacting regime ($\gamma\ll 1$) we
have $\mu=ng$, and in the strongly intertacting Tonks-Girardeau
regime the chemical potential is
\begin{equation}   \label{muLLlarge}
\mu=\frac{\pi^2\hbar^2 n^2}{2m}\left(1-\frac{8}{3\gamma}\right) \quad ;
\qquad \gamma\gg 1 \quad .
\end{equation}

In the presence of an (axial) trapping potential $V(z)=m\omega^2z^2/2$,
one should add this potential to the Hamiltonian (\ref{LLham}). Then, for an 
arbitrary $\gamma$ the model is no longer integrable. However, if the chemical
potential of the 1D trapped gas satisfies the inequality $\mu_0\gg\hbar\omega$, 
one can use the local 
density approximation \cite{Dunko,Ciara2}. Namely, assuming local equilibrium, the density
distribution is governed by the equation
\begin{equation}    \label{ld}
\mu(n(z))+V(z)=\mu_0 \quad ,
\end{equation}
where the local value of the chemical potential $\mu(n(z))$ follows from the
solution of the Lieb-Liniger model for a uniform gas with density equal to
$n(z)$. In other words, the local chemical potential is given by
Eq.(\ref{muLL}) in which $n=n(z)$ and $\gamma=mg/\hbar^2n(z)$.

As $\mu(n(z))$ should be positive at any local density, Eq.(\ref{ld}) leads to the
Thomas-Fermi density profile with a (half)size $R_{TF}=(2\mu_0/m\omega^2)^{1/2}$.
So, in the interval of distances, $-R_{TF}<z<R_{TF}$, the density distribution
is governed by Eq.(\ref{ld}), and otherwise one has $n(z)=0$. The density is
maximum at the trap center ($z=0$) and it smoothly decreases to zero when
moving to the Thomas-Fermi border of the trapped gas. The normalization
condition 
\[
\int_{-R_{TF}}^{R_{TF}}\!\!dz\, n(z) = N \quad ,
\]
gives a relation between the chemical potential $\mu_0$ and the number of
particles $N$.

\subsection{Phase coherence at zero temperature}

In Lecture 3 we showed that for a realistic (axial) size of the  
sample, the ground state of the weakly interacting trapped 1D Bose  
gas ($\gamma\ll 1$ and $T=0$) is a true Bose-Einstein condensate. 
The question now is how deeply one should be in the weakly interacting 
regime that this statement is valid. In other words, how small should 
be the parameter $\gamma$ at the trap center for having the full phase 
coherence of the trapped 1D gas. 

We will consider a harmonically trapped 1D Bose gas in the Thomas-Fermi regime
at $T=0$ and calculate the single-particle correlation function $g_1(z,z')$
for distances $|z-z'|$ greatly exceeding the correlation length $l_c$ \cite{Gangardt1}.
Assuming {\it a priori} small density fluctuations at such distances, we will
rely on the 1D hydrodynamic approach \cite{Haldane} which describes long-wave
properties of the 1D fluid in terms of the density fluctuations $\delta\hat
n$ and phase $\hat\phi$. Small fluctuations of the density at large
distances lead to linearized continuity and Euler equations:
\begin{eqnarray}       
&&\pi\frac{\partial\delta\hat n}{\partial t}=-\frac{\partial}{\partial
z}v_J(z)\frac{\partial\phi}{\partial z} \quad , \label{Lutn}     \\ 
&&\frac{\partial\phi}{\partial t}=-\pi v_N(z)\delta\hat n(z) \quad ,  \label{Lutphi}
\end{eqnarray}
where the velocities $v_J$ and $v_N$ are given by
\begin{eqnarray}        
&&v_J=\frac{\pi\hbar n(z)}{m} \quad ,     \label{vJ}  \\
&&v_N=\frac{1}{\pi\hbar}\frac{\partial\mu}{\partial n}\Big\vert_{n=n(z)} \quad . 
\label{vN} \end{eqnarray}
and the ratio $K(z)=\sqrt{v_J(z)/v_N(z)}$ is the local value of the 
Luttinger parameter.
Using commutation relations (\ref{overview:comrelphase}) equations (\ref{Lutn}) 
and (\ref{Lutphi})
can be obtained as equations of motion from the quantum hydrodynamic
Hamiltonian:
\begin{equation}       \label{Haldaneham}
\hat H=\frac{\hbar}{2\pi}\int\!\!dz\,
\left(v_N(z)(\pi\delta\hat
n)^2+v_{J}(z)(\partial_z\hat\phi)^2\right)
=\sum_j\varepsilon_j\hat b^{\dagger}_j\hat b_j \quad ,
\end{equation}
where $\varepsilon_j$ and $\hat b_j$ are eigenenergies and annihilation
operators of elementary excitations. Note that in the uniform case these
excitations are phonons and the velocity of sound is $c_s=\sqrt{v_Jv_N}$.
Similarly, for the trapped gas the local value of the sound velocity is
$c_s=\sqrt{v_J(z)v_N(z)}$. The Hamiltonian (\ref{Haldaneham}) is a
generalization of the effective Hamiltonian of Haldane \cite{Haldane} to a 
non-uniform system.

The solution of Eqs.~(\ref{Lutn}) and (\ref{Lutphi}) is given by the expansion
of operators $\delta\hat n$ and $\phi$ in eigenmodes characterized by an
integer quantum number $j>0$:
\begin{eqnarray}
&&\delta\hat n(z,t)=\sum_j\left(\frac{\varepsilon_j}{2\pi\hbar v_N(0)R_{TF}}
\right)^{1/2}f_j(z)\hat b_j\exp{(-i\varepsilon_jt/\hbar)}+ {\rm h.c.}  \quad , \label{deltanu}  \\
&&\hat\phi(z,t)=\sum_j -i\left(\frac{\pi\hbar
v_N(0)}{2\varepsilon_jR_{TF}}\right)^{1/2}f_j(z) \hat
b_j^{\dagger}\exp{(-i\varepsilon_jt/\hbar)}+ {\rm h.c.} \quad . \label{deltaphiLut}
\end{eqnarray}  
The eigenfunctions $f_j(z)$ are normalized by the condition
\[
\int_{-1}^{1}\!\!dx\,\frac{v_N(0)}{v_N(x)} f^*_j(x)f_{j'}(x) = \delta_{jj'} \quad ,
\]
where we have introduced a dimensionless coordinate $x=z/R_{TF}$. Equations 
(\ref{Lutn}) and (\ref{Lutphi}) are then reduced to the eigenmode equation:
\begin{equation}         \label{eigenmodeLut}
v_N(z)\frac{\partial f(z)}{\partial z}v_J(z)\frac{\partial f(z)}{\partial z}
+\frac{\varepsilon_j^2}{\hbar^2}f_j(z)=0 \quad .
\end{equation}
Considering $f_j$ as a
function of the reduced coordinate $x=z/R_{TF}$, Eq.(\ref{eigenmodeLut})
takes the form:  
\begin{equation}        \label{eigenmode}
(1-x^2)f''_j-(2x/\beta(x))f'_j+(2\varepsilon^2/\beta(x)\hbar^2\omega^2) f_j = 0 \quad .
\end{equation}
The quantity $\beta(x)=d\ln\mu/d\ln n$ is determined by the local 
parameter $\gamma(z)=mg/\hbar^2n(z)$. In the Tonks regime
($\gamma\rightarrow\infty$) we have $\beta=2$, and in the weakly interacting
regime $\beta=1$. 

The coordinate dependence of $\beta$ is smooth and we
simplify Eq.~(\ref{eigenmode}) by setting $\beta(z)=\beta
(\gamma_0)\equiv\beta_0$, where $\gamma_0$ is the value of $\gamma$ at maximum
density.  This simplification has been recently used
\cite{Ciara2,Combescot} to study the excitation spectrum of trapped
1D Bose gases. Then Eq.~(\ref{eigenmode}) 
yields the spectrum 
\[
\varepsilon_j^2=\hbar^2\omega^2 (j\beta_0/2)(j+2/\beta_0-1) \quad ,
\] 
and the eigenfunctions $f_j (x)$ are expressed through Jacobi polynomials:
\[
f_J(x)=\frac{\sqrt{(j+\alpha+1/2)\Gamma(j+1)\Gamma(j+2\alpha+1)}}
{2^{\alpha}\Gamma(j+\alpha+1)}P^{(\alpha,\alpha)}_j(x) \quad ,
\]
where $\alpha=1/\beta_0-1$. 

As the density fluctuations on a large distance scale are small, for the
single-particle correlation function one can write
\begin{equation}      \label{phicorLut}
g_1(z,z')\equiv \langle\hat\Psi^{\dagger}(z)\hat\Psi(z')\rangle =
\sqrt{n(z)n(z')}\exp\left(-\frac{1}{2}\langle
(\hat\phi(z)-\hat\phi(z'))^2\rangle \right) \quad . 
\end{equation}
Using Eq.~(\ref{deltaphiLut}) the mean square fluctuations of the phase in the
exponent of Eq.~(\ref{phicorLut}) are reduced to the sum over $j$-dependent
terms containing eigenfunctions $f_j$ and eigenenergies
$\varepsilon_j$.  For the vacuum phase fluctuations this sum is
logarithmically divergent at large $j$, which is similar to the high-momentum
divergence in the uniform case.
Accordingly, we introduce a cut-off $j_{max}$
following from the condition $\varepsilon_j \approx \min\{\mu(z),\mu(z')\}$ 
and ensuring a 
phonon-like character of excitations at distances $z$ and $z'$.  
This is equivalent to having only the low-energy part of the phase fluctuations,
$\langle(\hat\phi_p(z)-\hat\phi_p(z'))^2\rangle$, in the exponent of 
Eq.(\ref{phicorLut}) when considering the weakly interacting regime. 

The vacuum phase fluctuations have been calculated by using two
approaches: numerical summation over the eigenmodes with exact
$f_j,\varepsilon_j$ from the simplified Eq.~(\ref{eigenmode}), and
quasiclassical approach assuming that the main contribution comes from
excitations with $j\gg 1$.  In the latter case, for $z'=-z$ we
obtain  
\begin{equation}        \label{Haldres}
\left\langle\left(\hat\phi(z)-\hat\phi(-z)\right)^2\right\rangle\approx K^{-1}
(z) \ln\!\left( |2z| \over l_c (z)\right) \quad , 
\end{equation}
which is close to Haldane's result for a uniform system
\cite{Haldane} with the Luttinger parameter $K(z)$ and
correlation length $l_c(z)$. In the Tonks regime we have $K=1$ and 
$l_c=1/n(x)=1/n(0)\sqrt{1-(x/R_{TF})^2}$. Remarkably, in this limit
the hydrodynamical expression (\ref{Haldres}) coincides with 
the recently obtained exact result \cite{Forrester} for the single-particle correlation
function of a harmonically trapped Tonks gas.   
\begin{figure}
  \begin{center}
    \psfrag{y}{$y=z/R_{TF}$} \psfrag{g}{$g_1(z,-z)/n(z)$}
    \psfrag{g01}{$\gamma_0=0.1$} \psfrag{g1}{$\gamma_0=1$}
\psfrag{g10}{$\gamma_0=10$} \includegraphics[width=3.375in]{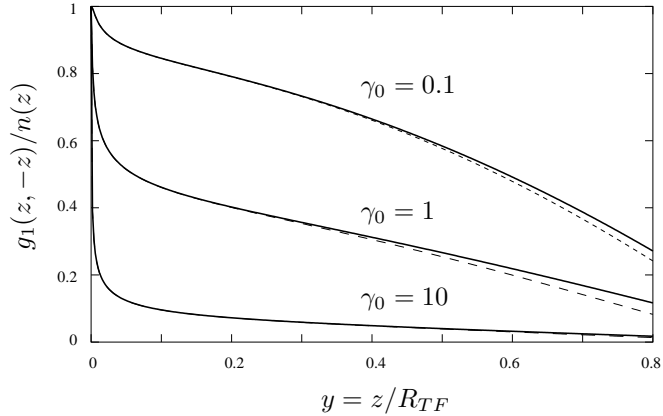}
    \caption{Normalized single-particle correlation function $g_1(z,-z)/n(z)$ 
versus $z/R_{TF}$, for
$N=10^4$ and various values of $\gamma_0$. The solid curves show
numerical results, and the dashed curves the results of the quasiclassical
approach.} 
    \label{fig:phasecorr.1}
  \end{center}
\end{figure}

The dependence of $g_1$ on the dimensionless
coordinate $x$ is governed by two parameters: $\gamma_0$ and the number of
particles $N$. In Fig.~\ref{fig:phasecorr.1} we present the quantity
$g_1(z,-z)/n(z)$ for $N=10^4$ and various values of $\gamma_0$. 
As expected, the phase coherence is completely lost in the strongly interacting
regime ($\gamma_0\gg 1$). Moreover, on a distance scale $z\sim R_{TF}$ the 
coherence is already lost for $\gamma_0\approx 1$, and the full phase coherence
requires $\gamma_0$ well below $0.1$.

\subsection{Local correlations at $T=0$}

The strong transverse confinement required for the 1D regime 
can lead to high 3D densities of a trapped gas. At a large number
of particles the 3D density can exceed $10^{15}$~cm$^{-3}$ and one expects
a fast decay due to 3-body recombination. It is then
crucial to understand how the correlation properties of the gas influence the
decay rate. For this purpose, we calculate local correlations in the 1D Bose
gas and show that the decay rates are suppressed  in the
Tonks-Girardeau and intermediate regimes, which is
promising for achieving these regimes with a large number of particles. 

The rate of 3-body recombination is proportional to the local 3-particle
correlation function $g_3=\langle\hat\Psi^\dagger(z)\hat\Psi^\dagger(z)
\hat\Psi^\dagger(z)\hat\Psi(z)\hat\Psi(z)\hat\Psi(z)\rangle$ \cite{Kagan85},
where all field operators are taken for the same time.
Similarly, the rates of 2-body inelastic processes involve the correlation
function $g_2
=\langle\hat\Psi^\dagger(z)\hat\Psi^\dagger(z)\hat\Psi(z)\hat\Psi(z)\rangle$. 
Assuming that local correlation properties are insensitive to the geometry of
the system we consider a uniform 1D gas of $N$ bosons on a ring of
circumference $L$. The gas is described by the Lieb-Liniger Hamiltonian
(\ref{LLham}) which we now rewrite in the second quantization: 
\begin{equation} \label{LLham2}
\hat H=\int\!\!dz
\left(\frac{\hbar^2}{2m}\frac{\partial\hat\Psi^{\dagger}}{\partial
z}\frac{\partial\hat\Psi}{\partial z}+\frac{g}{2}\hat\Psi^{\dagger}\hat
\Psi^{\dagger}\hat\Psi\hat\Psi\right) \quad .
\end{equation}   

For finding $g_2$ at $T=0$, we use the Hellmann-Feynman
theorem \cite{Hellmann1933,Feynman1939}. Namely, one shows that the
expectation value of the four-operator term in the Hamiltonian (\ref{LLham2})
is proportional to the derivative of the ground state energy with respect to
the coupling constant: 
\[
\frac{d E_0}{d g} = \langle \Phi_0 | \frac{dH}{dg} |\Phi_0\rangle =
\frac{g_2L}{2} \quad .
\] 
The first identity follows from the normalization of the ground
state wavefunction $\Phi_0$, and the second one is obtained straightforwardly
from the Hamiltonian (\ref{LLham2}). The
ground state energy is given by Eq.(\ref{E0}), and for the 2-particle
local correlation, we then obtain
\begin{equation}        \label{g2} 
   g_2 (\gamma) = n^2 \frac{d e(\gamma)}{d\gamma} \quad .
\end{equation}
In fact, the original work of Lieb and Liniger contains a similar calculation
of the interaction energy.

\begin{figure}
  \begin{center}
    \psfrag{xlabel}{$\gamma$} \psfrag{ylabel}{$g_2(\gamma)/n^2$}
    \includegraphics[width=3.375in]{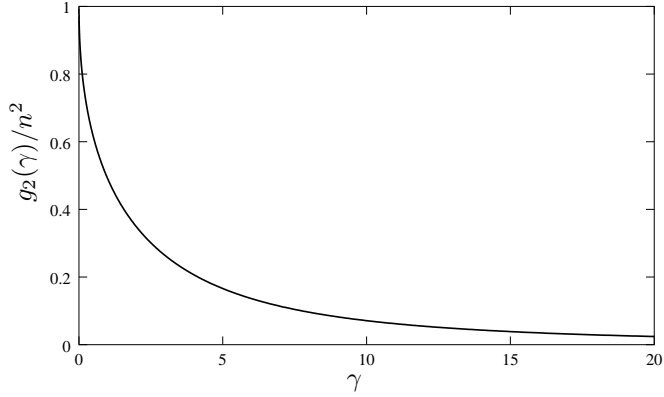}
    \caption{Local correlation function $g_2$ versus $\gamma$.}
    \label{fig:phaseterm.1}
  \end{center}
\end{figure}

The function $g_2 (\gamma)/n^2$ calculated by using numerical results for
$e(\gamma)$ from \cite{Ciara1}, is shown in Fig.~\ref{fig:phaseterm.1}.  
The quantum Monte Carlo calculations of $g_2$ \cite{Giorgini} arrive at the same results.
For small and large values of
$\gamma$, relying on Eqs.~(\ref{eweak}) and (\ref{estrong}), we obtain:
\begin{eqnarray} 
&&\frac{g_2(\gamma)}{n^2}= 1-\frac{2}{\pi}\sqrt{\gamma} \quad , \qquad \gamma\ll 1 \quad ;\label{g2weak}\\
&&\frac{g_2(\gamma)}{n^2}= \frac{4\pi^2}{3\gamma^2} \quad ,\qquad 
   \gamma\gg 1\quad .  \label{g2strong}
\end{eqnarray}

The results in Fig.~\ref{fig:phaseterm.1} and Eq.~(\ref{g2strong}) clearly
show that 2-particle correlations and, hence, the rates of pair inelastic
processes are suppressed for $\gamma \gtrsim 1$. This provides a possibility
for identifying the Tonks-Girardeau and intermediate regimes of a trapped 1D
Bose gas through the measurement of photoassociation in pair interatomic
collisions.

Note that for the weakly interacting regime the result of Eq.(\ref{g2weak})
can be obtained directly from the Bogoliubov approach. Writing the field
operator in the form (\ref{overview:dens-phase}) and using the commutation
relations ( \ref{overview:comrelphase}),
we find
$$
\hat\Psi^{\dagger}(z)\hat\Psi^{\dagger}(z')\hat\Psi(z)\hat\Psi(z')=
\hat n(z)\hat n(z')-\hat n\delta(z-z') \quad .
$$
We then represent the operator of the density as
$\hat n(z)=n+\delta\hat n(z)$ and confine ourselves to the second order in small
density fluctuations $\delta\hat n(z)$ around the mean density $n$. This gives 
$$
\langle\hat\Psi^{\dagger}(z)\hat\Psi^{\dagger}(z')\hat\Psi(z)\hat\Psi(z')
\rangle =n^2+\langle(\delta\hat n(z)\delta\hat n(z'))\rangle -n\delta(z-z') \quad .
$$
Using the expansion of the density fluctuations in terms of Bogoliubov
excitations, given by Eq.(\ref{overview:expansiondens}), we obtain: 
$$
\langle\hat\Psi^{\dagger}(z)\hat\Psi^{\dagger}(z')\hat\Psi(z)\hat\Psi(z')
\rangle=n^2+n\int_{-\infty}^{\infty}\!\frac{dk}{2\pi}\,\left[\,
|f_k^-|^2(1+2N_k)-1 \right] \exp\!\left[ik(z-z')\right]
\quad ,
$$
where $E_k=\hbar^2k^2/2m$, $\varepsilon_k=\sqrt{E_k^2+2\mu E_k}$ is the
Bogoliubov excitation energy, and the Bogoliubov function $f_k^-$ is determined
by Eq.(\ref{fpmun}). Then, taking the limit $z'\rightarrow z$ we arrive at the expression
for the local 2-body correlation:
\begin{equation}      
\label{g2Bog}
g_2=\langle\hat\Psi^{\dagger}(z)\hat\Psi^{\dagger}(z)\hat\Psi(z)\hat\Psi(z)
\rangle =n^2+n\int_{-\infty}^{\infty}\!\frac{dk}{2\pi}\, \left[\frac{E_k}
{\varepsilon_k}(1+2N_k)-1\right] \quad .
\end{equation} 
For $T=0$ the occupation numbers for the
excitations, $N_k=0$, and the integration of Eq.(\ref{g2Bog}) immediately gives the
result of Eq.(\ref{g2weak}).  

The 3-particle local correlation $g_3$ cannot be obtained from the
Hellmann-Feynman theorem. In the weakly interacting regime
($\gamma\ll 1$) one can use the Bogoliubov approach, which gives
Eq.(\ref{g2Bog}) with an extra factor 3 in the second term of the rhs.
At $T=0$ we then have
\begin{equation}
  \label{g3weak}
 \frac{g_3 (\gamma)}{n^3} \simeq 1-\frac{6}{\pi} \sqrt{\gamma}\quad , \qquad \gamma\ll 1 \quad .
\end{equation}

For the Tonks-Girardeau regime ($\gamma\gg 1$) we demonstrate a method
for calculating the leading behavior of local correlations. Details 
are given in \cite{Gangardt3}, and here we present a compact derivation
of $g_3$ at $T=0$ \cite{Gangardt1}. In the first quantization
the expression for this function reads
\begin{eqnarray}
\label{eq:g3first}
g_3(\gamma)=\frac{N!}{3!(N-3)!}\int\!\!dz_4\ldots
dz_N\,\left|\Phi^{(\gamma)}_0(0,0,0,z_4,\ldots,z_N)\right|^2 \quad ,
\end{eqnarray}
where $\Phi^{(\gamma)}_0$ is the ground state function given in the domain
$0<z_1<\ldots<z_N<L$ by the Bethe Ansatz solution:
\begin{eqnarray}
\Phi^{(\gamma)}_0(z_1,z_2,\ldots,z_N)\propto
\sum_{P}a(P)\exp\left(i\sum k_{P_j}z_j\right) \quad ,
\label{eq:gsfunc}
\end{eqnarray}
where $P$ is a permutation of $N$ numbers, quasimomenta $k_j$ are solutions
of the Bethe Ansatz equations, and
\begin{eqnarray}
a(P) = \prod_{j<l} \left(\frac{i\gamma n + k_{P_j}-k_{P_l}}
{i\gamma n - k_{P_j}+k_{P_l}}\right)^\frac{1}{2} \quad .  \nonumber
\end{eqnarray}
For $\gamma\gg 1$, we extract the leading contribution to $\Phi_0^{(\gamma)}$
at three coinciding points by symmetrizing the amplitudes $a(P)$ over the first
three elements of the permutation $P$:
\begin{eqnarray}
\label{eq:apsym}
\frac{1}{3!}\sum_p  a( P_{p_1}, P_{p_2}, P_{p_3},P_4,
\ldots P_N)\!\simeq
\frac{\varepsilon_P}{(i\gamma n)^3}
\prod_{j<l}(k_{P_j}-k_{P_l}) \quad ,    
\end{eqnarray} 
where $j,l=1,2,3$. The sign of the permutation $P$ is $\varepsilon_P$, and $p$
runs over six permutations of $1,2,3$. For large $\gamma$, the difference of
quasi-momenta $k_j$ from their values at
$\gamma=\infty$  is of order $1/\gamma$ and can be neglected. 
Then, from Eqs.~(\ref{eq:gsfunc}) and (\ref{eq:apsym}) we
conclude that to this level of accuracy the ground state wave function at
three coinciding points is given by derivatives of the wave function of free
fermions $\Phi^{(\infty)}_0(z_1,z_2,z_3,z_4,\ldots)$ at $z_1=z_2=z_3=0$:
\begin{eqnarray}
\label{eq:gslead}
\Phi^{\!(\gamma)}_0(0,0,0,z_4,\ldots)\simeq
-\frac{1}
{(\gamma n)^3} 
\Big[\prod_{j<l}\!(\partial_{z_j}-\partial_{z_l})\Big]\Phi^{(\infty)}_0 \quad .
\end{eqnarray}
Substituting Eq.~(\ref{eq:gslead}) into Eq.~(\ref{eq:g3first}) we
express the local correlation $g_3$ through derivatives of the 3-body
correlation  function of free fermions. Using  Wick's theorem the
latter  is given by  a sum of products of 1-particle fermionic Green
functions $G(x-y)= \int_{-k_F}^{k_F}\!dk\, e^{ik(x-y)}/2\pi$, where $k_F=\pi n$
is the Fermi wavevector. The calculation from
Eq.~(\ref{eq:g3first}) is then straightforward  and 
in the considered limit of $\gamma\gg 1$ we obtain
\begin{equation}
\label{eq:g3final}
\frac{g_3(\gamma)}{n^3}=\frac{36}{\gamma^6 n^9}\left[\left(G''
(0)\right)^3-G^{(4)}(0)G''(0) G(0)\right]\quad ,  
\end{equation}
which yields
\begin{equation}     \label{g3strong} 
\frac{g_3}{n^3}=\frac{16\pi^6}{15\gamma^6}\quad .
\end{equation}

This result for $g_3$ and the result of Eq.(\ref{g2strong}) for $g_2$, have a 
transparent physical explanation. A characteristic distance related to the
interaction between particles is $r_{g}=\hbar ^{2}/mg\sim 1/\gamma n$, and
the strong repulsion between particles provides fermionic correlations at 
interparticle distances $z\gtrsim r_{g}$.
For smaller $z$ the correlation functions practically do not change.
Therefore, the local correlation $g_2$ at a finite large
$\gamma $ is nothing else than the pair correlation function for
free fermions at a distance $r_{g}$. The latter is $g_2\sim
n^2(k_{F}r_{g})^{2}\sim n^2/\gamma ^{2}$,
which agrees with the result of Eq.(\ref{g2strong}).
Similarly, $g_3$ is the free-fermion 3-particle correlation function at distances 
$\sim r_g$. It is approximately equal to the product of three pair correlation
functions, i.e. we have $g_3\sim n^6/\gamma ^{6}$ in qualitative
agreement with Eq.(\ref{g2strong}).

Thus, from Eq.~(\ref{g3strong}) we conclude that the 3-body decay of 1D
trapped Bose gases is strongly suppressed in the Tonks-Girardeau regime.
Moreover, Eq.~(\ref{g3weak}) shows that even in the weakly interacting
regime of a quasicondensate, with $\gamma\approx 10^{-2}$, one has a 20\%
reduction of the 3-body rate. Thus, one also expects a significant reduction
of the 3-body decay in the intermediate regime.

For $l_0\gg a$, the 3-body recombination process in 1D trapped gases occurs at
interparticle distances much smaller than $l_0$. Therefore, the equation for
the recombination rate is the same as in 3D cylindrical Bose-Einstein
condensates with the Gaussian radial density profile $n_{3D}=(n/\pi l_0^2)
\exp \left(-\rho^2/l_0^2\right)$, where $\rho$ is the radial coordinate. There
is only an extra reduction by a factor of $g_3/n^3$. A characteristic decay
time $\tau$ is then given by the relation
\begin{equation}
\label{eq:tau}
\frac{n}{\tau}=g_3\alpha_{3D}\int\! 2\pi\rho d\rho\, \left(\frac{n_{3D}}{n}\right)^3=
\frac{\alpha_{3D}g_3}{3(\pi l_0^2)^2}\quad ,
\end{equation}
where $\alpha_{3D}$ is the recombination rate constant for a 3D 
condensate. Even
for 3D densities $n/\pi l_0^2\!\sim\!10^{15}$ cm$^{-3}$ the life-time 
$\tau$ can
greatly exceed seconds when approaching the Tonks-Girardeau regime. 

In the recent rubidium experiment at NIST \cite{Phi03}, an array of 1D tubes of bosons
was created by optically confining the atoms with a radial frequency
$\omega_0\approx 30$ kHz ($l_0\approx 350$ \AA). Then, for the number of atoms
$N\approx 200$ in each 1D tube, the intermediate regime with $\gamma\sim 1$ has
been achieved. The measurement of the 3-body decay showed a reduction of the
rate by approximately a factor of seven.

\subsection{Finite-temperature local correlations}

We now discuss local correlations in 1D Bose gases at finite temperatures
and confine ourselves to two-body correlations in the uniform case \cite{Gangardt2}.
The calculation of $g_2$ will allow us to identify the various regimes of
quantum degeneracy, which is important for the development of possible atom
lasers in 1D waveguides.
At zero temperature there are two physically distinct
regimes of quantum degeneracy: the weakly interacting regime of a
quasicondensate for $\gamma\ll 1$, and the strongly interacting Tonks-Girardeau
regime where the 1D gas undergoes ``fermionization''.
These regimes are also present at finite temperatures. However, we will
see that for a very small interaction strength one can have a decoherent
quantum regime, where the fluctuations are enhanced and reach 
the non-interacting Bose gas level with $g_2\rightarrow 2n^2$ (rather
than $g_2\rightarrow n^2$ characteristic for the quasicondensate regime).
We will also see that for a large interaction strength the reduction of
local correlations is present at temperatures exceeding the temperature of
quantum degeneracy $T_d=2\pi\hbar^2n^2/m$ and, in this respect, one has the
regime of "high-temperature fermionization".

So, we again consider a 1D uniform gas of $N$ bosons on a ring of circumference
$L$, described by the Hamiltonian (\ref{LLham}) or, equivalently, by the
Hamiltonian (\ref{LLham2}). The exact solution of this problem at finite $T$
has been found by Yang and Yang \cite{YY} using the Bethe Ansatz. They derived
exact integral equations for thermodynamic functions and proved their
analyticity. Note that in trapped gases at finite temperatures, the 1D regime
requires the thermal de Broglie wavelength of particles $\Lambda_T$ to be
much smaller than the amplitude of zero point oscillations in the tightly
confined directions, $l_0$. This requirement should be added to
Eq.(\ref{ineq}), and we thus obtain the inequalities 
$a\ll l_{0}\ll \{1/n,\, \Lambda _{T}\}$. They allow us to
analyze finite-temperature correlation properties of the 1D trapped gas 
at any $\gamma$ by using the Yang-Yang results.

The behavior of the 2-body local correlation at finite temperatures is
governed by two parameters: $\gamma $ and $\tau=4\pi T/T_d$. We first show how
$g_2$  has been calculated by using the Hellmann-Feynman theorem 
\cite{Hellmann1933,Feynman1939}. Consider the partition function $Z=\exp
(-F/T)=\mathrm{Tr}\exp (-\hat H/T)$ which determines the free energy $F$. Here
the trace is taken over the states of the system with a fixed number of
particles in the canonical formalism or, for the grand canonical description,
one has to replace the condition of a constant particle number by the
condition of a constant chemical potential $\mu $ and add the term $-\mu \hat
N$ to the Hamiltonian. For the derivative of the free energy with respect
to the coupling constant one has 
\begin{eqnarray}
\frac{\partial F}{\partial g}=-T\frac{\partial \log Z}{\partial g}
=\frac{1}{Z}{\textrm{Tr}}\left[ \frac{\partial \hat H}{\partial g}
\exp\!\left(-\frac{\hat H}{T}\right) \right]=\frac{g_2L}{2}\quad .
\label{eq:hellfeyn} 
\end{eqnarray}
Introducing the free energy per particle $f(\gamma ,\tau )=F/N$,
the 2-particle correlation function is:
\begin{equation}
g_2=\langle \Psi ^{\dagger }\Psi ^{\dagger }
\Psi \Psi \rangle=\frac{2m}{\hbar ^2}\left(\frac{\partial
f(\gamma ,\tau )}{\partial \gamma }\right)_{n,\tau }\quad .\label{gnorm}
\end{equation}
In Fig.~12 we present the results for $g_2$, found by numerically
calculating $f(\gamma,\tau)$ from the Yang-Yang exact integral equations
and then using Eq.(\ref{gnorm}) \cite{Gangardt2}.  

\begin{figure}
\begin{center}
\includegraphics[  width=8cm]{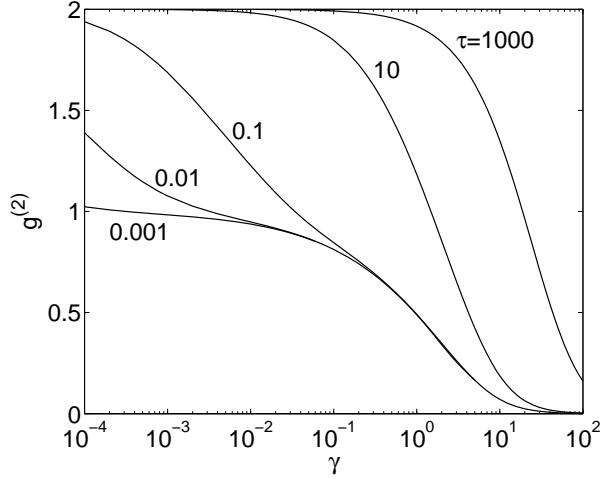}
\caption{Correlation function $g^{(2)}=g_2/n^2$ versus $\gamma $ at different
$\tau=4\pi T/T_d $.}
\end{center}
\label{g2vsgamma_log}
\end{figure}

We now give a physical description of different regimes determined
by the values of the coupling constant $\gamma $ and the reduced
temperature $\tau $ \cite{Gangardt2}. As well as at $T=0$, 
in the strongly interacting
regime the correlation function $g_2$ reduces dramatically due to the
strong repulsion between particles, and the physics resembles that of free
fermions. Interestingly, this regime is realized both below and above the
quantum degeneracy temperature. Similarly to $g_3$ and $g_2$ at zero
temperature, the finite-temperature $g_2$ can be expressed through
derivatives of the Green function of free fermions $G(x)=\int\!dk\,n_{F}(k)\exp
{(ikx)}/(2\pi )$, where $n_{F}(k)$ are occupation numbers for free fermions.
For the correlation function we obtain
$$
\frac{g_2}{n^2}=\frac{4}{\gamma ^{2}n^{4}} \left[\left(G^{\prime }(0)\right)^{2}-G^{\prime \prime
}(0)G(0)\right] \quad .
$$

In the regime of quantum degeneracy, $\tau \ll 1$, the local correlation
is dominated by the ground state distribution $n_{F}(k)=\theta
(k_{F}^{2}-k^{2})$,
where $k_{F}=\pi n$ is the Fermi momentum, and one only obtains a small 
finite-temperature correction to the zero-temperature result 
(\ref{g2strong}).

In the temperature interval $1\ll \tau \ll \gamma ^{2}$ the gas is
non-degenerate, but the interaction length $r_{g}$ is still much smaller
than the thermal de Broglie wavelength $\Lambda _{T}$. As well as at zero temperature,
$g_2$ can be viewed as the pair correlation function for free fermions at a 
distance $\sim r_g$. Taking into account that the characteristic momentum of 
particles is now the thermal momentum $k_{T}=\sqrt{2mT}/\hbar $, one finds 
$g_2\sim n^2(k_{T}r_{g})^{2}\sim n^2\tau /\gamma ^{2}$.
Calculating the Green function $G(x)$ for the classical distribution
$n_{F}(k)$, we obtain: 
\begin{equation}
\frac{g_2}{n^2}=\frac{2n^2\tau}{\gamma ^{2}} \quad , \qquad  1\ll \tau \ll \gamma
^{2} \quad ,
\label{eq:III}
\end{equation}
which agrees with the given qualitative estimate. The correlation
function $g_2/n^2$ is still much smaller than unity and we thus have
a regime of ``high-temperature fermionization''.
The result of Eq.~(\ref{eq:III}) agrees with
the outcome of numerical calculations. 

In the weakly interacting regime of a quasicondensate, 
one may use Eq.(\ref{g2Bog}). 
For temperatures $\tau \gg \gamma $, thermal fluctuations are more
important than vacuum fluctuations,
and we obtain: 
\begin{equation}
\frac{g_2}{n^2}=1+\frac{\tau}{2\sqrt{\gamma }}\quad  ,\qquad \gamma \ll \tau \ll \sqrt{\gamma}\ll
1 \quad .\label{eq:V}
\end{equation}
The phase coherence length is determined by thermal long-wavelength 
fluctuations of the phase. The calculation, similar to that for a trapped gas
in Lecture 3, gives $l_{\phi }\approx \hbar ^{2}n/mT$. The
condition $l_{\phi }\gg l_{c}$, which is necessary for the existence
of a quasicondensate and for the applicability of the Bogoliubov
approach, immediately yields the inequality $\tau \ll \sqrt{\gamma }$.
Thus, Eq. (\ref{eq:V}) is valid under the condition 
$\gamma \ll \tau \ll \sqrt{\gamma}$,
and the second term on the rhs of this equation is a small correction.
In the region of its validity, the result
of Eq.(\ref{eq:V}) agrees well with numerical data.  

At a very weak coupling strength given by
$\gamma \ll \tau^2\ll 1$, the gas is in a decoherent quantum
regime \cite{Castin,Gangardt2}, where both phase and density fluctuations 
are large
and the local correlation is always close to the result for free
bosons, $g_2=2n^2$. The only consequence of quantum degeneracy
is the quantum Bose distribution for occupation numbers of particles,
and the decoherent quantum regime can be treated asymptotically by
employing a standard perturbation theory with regard to the coupling
constant $g$.

The presence of the quantum decoherent regime helps to explain the apparent 
contradiction that in the thermodynamic limit a \emph{free} Bose gas at any finite
temperature is known to display large thermal (Gaussian) density fluctuations 
with $g_2=2n^2$. For the 3D gas this result requires the grand canonical
description \cite{ZiffUhlenbeckKac77}, whereas in 1D and 2D it is
valid for any choice of the ensemble. On the other hand, the widely used
mean-field Bogoliubov approach for an interacting Bose gas leads to
$g_2\approx n^2$. 
The above discussed results explain this fact in the exactly solvable
1D case: there is a \emph{continuous} transition from the quasicondensate to
decoherent regime, depending on the density and temperature. As $\gamma $
is decreased towards a free gas, the quasicondensate result of $g_2\simeq
n^2$ only holds above a certain interaction strength. Below this, there
is a dramatic increase in fluctuations, with $g_2\rightarrow 2n^2$
in the free gas limit.

{\it Problem}: Prove that the density fluctuations for the uniform Tonks-Girardeau 
gas are small in the limit of large distances. Find the distance scale on which they
become large.

\subsection{Acknowledgements}

These lectures have been given by Gora Shlyapnikov at the Les Houches summer school
"Quantum Gases in Low Dimensions", organized in 2003 by Ludovic Pricoupenko, 
H\'el\`ene Perrin,
and Maxim Olshanii. Dmitry Petrov and Gora Shlyapnikov are grateful to them for support 
and for hospitality at Les Houches. 

The authors acknowledge fruitful discussions with Carlos Lobo and Jook Walraven.
The work on these lectures was financially supported by the Nederlandse 
Organisatie voor Wetenschappelijk Onderzoek (NWO), by the Stichting voor Fundamenteel 
Onderzoek der Materie (FOM), by the French Minist\`ere de la Recherche et des 
Technologies, by INTAS, and by the Russian Foundation for Fundamental
Research. LKB est UMR 8552 du CNRS, de l'ENS et de l'Universit\'e Pierre et
Marie Curie. 


\end{document}